\documentclass[preprint2]{aastex}
\usepackage{graphicx}
\usepackage{xcolor}

\def\lsun{L$_{\odot}$\/\ }
\def\msun{M$_{\odot}$\/\ }

\slugcomment{Not to appear in Nonlearned J., 45.}

\shorttitle{radio-loud and optical AGN in different environments}
\shortauthors{H. Miraghaei }

\begin{document}


\title{The effect of environment on AGN activity: the properties of radio and optical AGN in void, isolated and group galaxies}

\author{Halime Miraghaei\altaffilmark{1}}
\altaffiltext{1}{Email:h.miraghaei@maragheh.ac.ir}

\email{Research Institute for Astronomy and Astrophysics of Maragha (RIAAM), 
University of Maragheh, Maragheh, Iran}

\begin{abstract}
The evolution of galaxies depends on their environments.
In this work, active galactic nucleus (AGN) activity in different environments 
has been studied. The fractions of radio and optical AGN
in four different environments have been compared using samples of void, isolated, 
group member, and the brightest group galaxies (BGGs). Galaxies in voids show significantly
lower stellar ages, concentrations, colours and surface mass densities, and they experience more
one-on-one interactions compared to the isolated galaxies
and galaxies in groups. In order to study pure environmental effects, the 
biases caused by the stellar mass and galaxy type quantified by 4000$\AA$ break 
strength have been removed. While the results confirm no dependence
of the optical AGN activity on environment in blue galaxies and with lower significance
in green galaxies, a higher fraction of optical AGN has been observed 
for the massive red galaxies in voids compared to the galaxies in dense environments.
This may be related to the higher amount of one-on-one interaction observed in the void galaxies, or it may reflect
more fundamental differences in the host galaxies or environments of the voids.
The radio-mode AGN activity increases in dense environment for red galaxies. 
No changes in the radio-loud AGN fraction have been
observed for the blue and green galaxies. This shows that the effect of environment
on AGN activity is not significant in the presence of cold gas in galaxies.
We also discuss whether the efficiency of gas accretion depends on the properties of the host galaxy.     

\end{abstract}


\section{Introduction}
\label{sec:Intro}

Supermassive black holes (SMBH) at the heart of massive galaxies play an essential
role in the evolution of galaxies and their environments. SMBHs with the substantial gas accretions
construct active galactic nuclei (AGN), which send huge amount of momentum and energy into the intergalactic medium (IGM) 
in the form of radiation, outflows and jets. The imprint of this activity is displayed in 
the entire electromagnetic spectrum from radio waves to gamma-rays and it is widely used for
the identification of AGN. The AGN can be selected via the detection of specific $\it{optical}$
emission lines as optical AGN or via the detection of powerful $\it{radio}$ jets as radio-loud AGN.

There are two major classifications for AGN. AGN are 
classified into radiatively efficient (quasar-mode) and 
radiatively inefficient (radio or jet-mode) based on their accretion rates into the SMBHs and are classified
into radio-loud and radio-quiet based on their  
radio luminosities. A significant fraction of the radio-loud AGN
population is radiatively inefficient, while radio-quiet AGN detected in optical wavelengths display 
two modes of accretion rates. The host galaxies of AGN in each 
classes are also different. Quasar-mode AGN activity is observed in blue star-forming galaxies
while radio-mode AGN activity is dominant in massive elliptical galaxies with old stellar populations.
The origin of the quasar/radio-mode dichotomy is well-explained by the 
fueling mechanism of the SMBH. The fresh cold gas in blue galaxies is accreted at a high rate 
in quasar-mode AGN while the huge reserviour of hot 
IGM gas in the environment of red elliptical galaxies, accreted slowly to the center, 
feeds radio-mode AGN (see Heckman \& Best 2014 for a review). 
In contrast, the origin of the radio-loud/radio-quiet dichotomy
is still unknown, but the black hole spin may be responsible 
(Garofalo et~al.\ 2010; McNamara et~al.\ 2011).

In addition to the host galaxy properties, the environments of AGN have also been studied
to find out a possible connection between AGN activity and 
 the close or large-scale environments. This has been explored for AGN samples selected 
in various electromagnetic wavelengths. In this regard, quite different results have been obtained,
variously reporting enhancement, decrease or no change in AGN activity at dense environments 
(Miller et~al.\ 2003; Kauffmann et~al.\ 2004; Gilmour et~al.\ 2007; 
Bradshaw et~al.\ 2011; Malavasi et~al.\ 2015; Manzer \& De Robertis 2014; Koulouridis et~al.\ 2018). 
The discrepancies in the results are likely to be because of 
fundamental differences in powering AGN at different wavelengths as discussed above,
differences in definitions of AGN activity (Man et~al.\ 2019) and environment or due to 
the possible biases in the samples of AGN which compare different environments.
The latter suggests these results need to be studied for different
types of galaxies, because AGN activity has been shown to be a strong function of 
host galaxy properties such as mass or colour (Best et~al.\ 2005b; Janssen et~al.\ 2012)
and host galaxy properties are strong functions of environment 
(Balogh et al. 2004; Baldry et~al.\ 2006).

The optical AGN activity for galaxies of different types has been investigated in some
recent works. Argudo-Fernandez et~al.\ 2018
showed an enhancement in the fraction of AGN with denser environments 
in quenched isolated galaxies. In a similar study, using samples of early and late-type galaxies, 
Lopes et~al.\ 2017 argued that AGN favored environments with lower relative velocities
such as fields, poor groups and cluster outskirts. In contrast, Sabater et~al.\ 2015
used a sample of galaxies matched in  
their masses and specific star formation rates (sSFRs) and showed that the effect of environment on the prevalence of 
optical AGN activity and luminosity is not significant.
All these studies show that the role of environment on optical AGN activity
 is still challenging to discern based on the literature.

Previous studies have shown that the radio-mode AGN activity strongly increases in dense environments 
(Best et~al.\ 2007; Sabater et~al.\ 2013; Malavasi et~al.\ 2015). 
This result is based on samples that include mostly red and quenched galaxies. 
Janssen et~al.\ (2012) studied galaxies of different colours
and show that ,although the radio AGN are dominant in red galaxies,
the presence of cold gas in blue galaxies enhances the AGN fraction.
Therefore, for radio-mode AGN activity, the effect of the environment needs to be
investigated for different types of galaxies.

In addition to the biases caused by the host galaxy types, the definition of
environment may also influence the results. 
Different environmental properties have been used to measure the level of
galaxy interaction such as:
local density or overdensity defined based on the nth nearest neighbour 
(Man et~al.\ 2019) or on the
projected two-point cross-correlation function (Wang \& Li 2018);
number of galaxies within a specific radius (Malavasi et~al.\ 2015); 
the level of tidal interaction (Sabater et~al.\ 2013);
galaxies in field, void galaxies and isolated galaxies versus galaxies in 
pairs, groups, clusters, filaments and walls ( von der Linden et~al.\ 2007; Manzer \& De Robertis 2014;
Argudo-Fernandez et~al.\ 2016, 2018; Magliocchetti et~al.\ 2018; Amiri  et~al.\ 2019); 
galaxy group richness or halo mass (Li et~al.\ 2019); the distance
to the center of galaxy group or galaxies in the core versus outskirt 
(Pimbblet et~al.\ 2013; Lopes et~al.\ 2017; Gordon et~al.\ 2018); 
group luminosity gap (Miraghaei et~al.\ 2014, 2015; Khosroshahi et~al.\ 2017) and
status as a merging or non-merging system (Hong et~al.\ 2015).

The main goal of this work is to study both the radio and optical AGN activity in 
different environments to find out how the interaction of host galaxy with the
environment affects the accretion into the central black hole. Samples of galaxies with different colours
are used to remove the bias caused by the galaxy type.  
The Sloan Digital Sky Survey seventh Data Release (SDSS DR7; York et~al.\ 2000) is used to construct 
four samples of galaxies in different environments: i) galaxies in voids, ii) isolated galaxies,
iii) galaxies in groups and iv) the brightest group galaxies. 
They are listed from the lowest to the highest levels of galaxy interaction.
Cosmological voids have the lowest environmental density in the radii up to tens of
Mpcs and the brightest group galaxies are normally located at the center
of groups and clusters as the most overdense regions formed in hierarchical galaxy formation.
A comparison between different definitions of environment 
has been performed using environmental properties defined in the 
literature and different environments used in this work.

The paper includes investigation of various optical properties
of galaxies at different environments including 
the cosmological voids as the most underdense regions in the Universe.
Previous studies have shown galaxies in voids to be fainter (Hoyle et~al.\ 2005), 
relatively bluer and more disk-like 
(Rojas et~al.\ 2004; Ricciardelli et~al.\ 2017) than non-void 
populations of galaxies. Kreckel et~al.\ (2012) show that void galaxies
with an ongoing gas accretion are still in the process of assembling. 
In contrast, galaxies in dense environments are preferentially early-type 
and red (Balogh et~al.\ 2004). These differences
are interpreted as a direct effect of environment on galaxy evolution. 
Galaxies in low-interaction environments, have shorter
merger histories and thus slower evolution.
We particularly distinguish between galaxies in voids and
isolated galaxies and make a comparison between these two types of noninteracting galaxies.
The optical AGN activity of noninteracting galaxies has also been
investigated by different authors. Amiri  et~al.\ 2019 show no enhancement in
the fraction of galaxies hosting optical AGN in voids compared to galaxy groups. 
Argudo-Fernandez et~al.\ 2016 show similar results
using samples of isolated and isolated pair galaxies. 
In contrast, Constantin et~al.\ 2007 show that optical AGN are more common in voids than walls.
In this study, both the optical AGN
activity and the radio-mode AGN activity in voids are investigated
 and the results have been compared with those for 
group galaxies and isolated galaxies.

The layout of this paper is as follows. The galaxy and radio source samples
 and the environmental parameters are presented in Section 2. 
Optical and radio AGN classifications are described in Section 3.
The overall properties of each galaxy sample are shown in Section 4.
The dependence of AGN activity on the stellar mass and colour is investigated in
Section 5. The radio and optical AGN activity are 
shown in Section 6.  Summary and conclusions
are presented in Section \ref{sec:summery}.  
Throughout the paper we assume
a $\Lambda CDM$ cosmology with the following parameters: $\Omega_m=0.3$,
$\Omega_\Lambda=0.7$ and $H_0=70$ km s$^{-1}$ Mpc$^{-1}$.\\

\section {Sample selection }
\label{sec:Sample}

The galaxy sample is based on the \textquoteleft main galaxy 
sample\textquoteright ~of the SDSS DR7 
(York et~al.\ 2000; Strauss et~al.\ 2002; Abazajian et~al.\ 2009). 
The optical spectra were drawn from
the value-added spectroscopic catalogues produced by the group
from the Max Planck Institute for Astrophysics and Johns Hopkins
University (MPA-JHU, cf. Brinchmann et al. 2004) which provides 818333 unique galaxy spectra
up to the redshift z=0.7.
Best $\&$ Heckman (2012) combined spectroscopic data from the \textquoteleft main galaxy
sample\textquoteright ~with radio data from the National Radio Astronomy 
Observatory Very Large Array (VLA) Sky Survey (NVSS; Condon
et al. 1998) and the Faint Images of the Radio Sky at Twenty 
centimetres (FIRST) survey (Becker, White $\&$ Helfand 1995) to derive
the largest SDSS radio catalogue of galaxies with their optical spectra (Best et~al.\ 2005a). 
This includes total stellar mass and black hole mass (Kauffmann et al. 2003b), 
4000$\AA$ break strength (stellar age),
galaxy magnitude, colour (rest-frame g-r),
 half-light radius as a galaxy size which is a radius containing 50$\%$ of the galaxy light
(R$_{50}$), concentration (C=R$_{90}$/R$_{50}$), 
half-light surface mass density:
$\mu_{50}$ = 0.5 M$_{\star}$/ ($\pi$
  R$_{50}$$^{2}$), 1.4 GHz total and core radio luminosity, 
radio size and morphological classifications
(see Miraghaei \& Best 2017 for the details).
The black hole mass has been calculated from the 
relation between the velocity dispersion (${\sigma}_{\star}$) of the galaxy 
and the black hole mass given in Tremaine
et~al.\ (2002): log(M$_{BH}$/\msun)=8.13 + 4.02 log(${\sigma}_{\star}$/200km s$^{-1}$).
Velocity dispersion estimates lower than about 70km s$^{-1}$ are not reliable given the
SDSS instrumental resolution for the spectra and the low signal to noise ratio in this range. 
This leads to a cut at log(M$_{BH}$/\msun)$\sim$6.3 for black hole masses in this study.

The SDSS group catalogue has been drawn from Tago et~al.\ 2010 
volume-limited group samples. Tago et~al.\ used a modified friends-of-friends (FOF) algorithm
to build five sets of galaxy groups with the various absolute magnitude limits 
for group members in the r-band, based on the SDSS magnitude limitation (r $<$ 17.77).
I adapted the subset of M$_{r}$ $\leq$ -20.0 (M$_{r}-$5log(h)$\leq$-20)
 which is a complete volume-limitted sample of galaxy groups up to 
the redshift, z=0.1. The catalogue provides isolated galaxies as galaxy groups with 
richness N=1. The galaxy group sample is defined by selecting those with N$\geq$2 in this study.

The sample of galaxies in voids is based on the Pan et~al.\ (2012) void catalogue. It 
provides 1054 statistically significant cosmic voids from the SDSS DR7 with radii $>$
10 h$^{-1}$ Mpc and a median effective radius of 17 h$^{-1}$ Mpc up to z=0.1.
There are 8046 galaxies located in voids, with an absolute magnitude 
limit of M$_{r}$=-20.0 or brighter, as it is required for this work. 
The void galaxies have been cross-matched
with the parent galaxy sample in this study using the SDSS unique Plate, MJD and Fiber IDs. 
The overlapping galaxies between the group/isolated samples 
and the void sample have been classified as void galaxies which were
593 group member galaxies including 393 BGGs and 7453 isolated galaxies. These were excluded
from the isolated, group member and BGG samples and were instead labeled as voids.
In total, the galaxy sample includes 8046 void galaxies, 61155 isolated galaxies, 15842 BGGs
and 46181 group member galaxies. The final number of galaxies in each class is listed in Table 1.

The environmental parameters introduced by Sabater et~al.\ (2013) have been added to this study. 
They are available for the SDSS DR7 main galaxy sample in the redshift range
0.03 $<$ z $<$ 0.1 that is compatible with the galaxy sample in this work. The primary parameters
are a) the local density defined as $$\eta = \log \left(\frac{3 k}{4 \pi r_{k}^{3}} \right),$$
where $r_{k}$ is the 
projected-distance (in Mpc) to the $k^{\mathrm{th}}$ nearest neighbour with k=10 and 
b) the tidal interaction defined as the relative tidal forces exerted by companions 
with respect to the internal binding forces of the target galaxy, and calculated using the equation
$$Q = 
\log \left( \sum_{i} \frac{Lr}{Lr_{i}} 
     \left( \frac{2 R}{d_{i}} \right)^{3} \right),$$ 
where $R$ is the estimated radius of the target galaxy, 
$d$ is the distance between the target and the companion (i) and
$Lr$ is the luminosity in the \textit{r}-band. In general, there is a correlation with a large
scatter between these two parameters which means a galaxy in a dense environment also experiences
a large tidal force. The correlation can be removed using principal component analysis (PCA), making
a new orthogonal set of parameters. Based on the PCA, Sabater et~al.\ (2013) use 
a combination of these parameters to define
PCA1 and PCA2 as a non-correlated set of environmental parameters that trace the overall interaction level
and one-on-one interaction respectively. PCA1 has positive contributions of density
and tidal interaction and PCA2 has negative contribution of density and positive contribution
of tidal interaction. Sabater et~al.\ (2013) also define a second set of
environmental properties besed on the PCA for group member galaxies using the richness parameter
provided by Tago et~al.\ (2010) as PCA1r, PCA2r and PCA3r
via combination of density, tidal interaction and richness.
Similar to the first set of parameters, PCA1r has positive contributions
of all three parameters, while PCA2r has positive contribution of tidal and negative
contributions of the other two parameters. PCA3r is composed of positive contribution of
richness and negative contributions of the rest. 
All of these sets, which are available publicly, are used in this study to quantify the level of
interaction in each set of galaxies.

\section{AGN classification}
\label{sec:AGNclass}

\begin{deluxetable}{lrrrr}
\footnotesize
\tablecaption{Number of sources in the sample of galaxies 
 with different environments  
\label{table1}}
\tablewidth{0pt}
\tablehead{
\colhead{} & \colhead{Galaxies} & \colhead{Isolated} &\colhead{Galaxies} &
\colhead{BGGs}    \\
\colhead{} & \colhead{in void} & \colhead{galaxies} &
\colhead{in group} &
\colhead{} }

\startdata
optical AGN&&&&\nl
 ~~\tablenotemark{a}total&2340 & 19648 &13274 &5514 \nl
~~\tablenotemark{b}L$_{\rm [OIII]}$$\geq$10$^{6.5}$ \lsun  &317 & 2394 &  1506 &649\nl
 radio AGN&&&&\nl
 ~~\tablenotemark{c}total&40 & 452 &800 &565 \nl
 ~~\tablenotemark{d}L$_{\rm NVSS}$$\geq$10$^{22.9} $ W/Hz &23 & 398&  762 &560 \nl
all galaxies \tablenotemark{e} &8046 & 61155 &  46181 & 15842 \nl
\enddata

\tablenotetext{a, b}{The number of optical AGN in each environments before and after applying [OIII] luminosity cut to the samples.} 
\tablenotetext{c, d}{The number of radio AGN in each environments before and after applying the radio luminosity cut to the samples.}

\tablenotetext{e}{The total number of galaxies including the AGN and 
normal galaxies in different environments used in this study.} 

\end{deluxetable}

AGN have been selected based on their radio power and optical emission lines
properties. This defines two samples of radio-mode and optical AGN
available in this study.

The optical AGN have been selected based on the BPT diagnostic diagram 
(Baldwin, Phillips $\&$ Terlevich 1981; Kauffmann et~al.\ 2003a)
using the ratio of optical emission lines. AGN, star forming (SF) galaxies and a 
population of transition or composite objects (SF/AGN) have 
been identified via this method (Kewley et al. 2006).
In this study, the optical AGN sample has been constructed by 
combining the objects identified as either AGN or transition. 
To have a complete sample of optical AGN up to z=0.1, the [OIII] emission line luminosity cut
at 10$^{6.5}$ \lsun has been applied to the galaxies selected as AGN (Sabater et~al.\ 2015).  
Therefore, most of the galaxies in the optical AGN sample are Seyfert galaxies.

The radio source classifications as SF or AGN have been drawn 
from Best $\&$ Heckman (2012). They adapted a combination of three different
methods based on i) 4000 $\AA$ break strengths and the ratio
of radio luminosity to stellar mass, ii) the ratio of 
radio-to-emission-line luminosity and iii) the BPT diagram 
to separate SF galaxies and the AGN.
To have a complete sample of radio AGN,
the 3mJy flux density limit of the NVSS
has been used to calculate the detection limit.
This corresponds to a 1.4 GHz luminosity of 
9$\times$10$^{22}$ W Hz$^{-1}$ at redshift z=0.1.
Therefore, the radio AGN sample is complete
to this limit up to the highest redshifts in this study.
The radio AGN with the total radio luminosity below this limit
are excluded from the radio-loud AGN sample. The number of optical and radio AGN
selected in this work in different environments are listed in Table 1.

Based on the AGN sample selection criteria described above, the 
radio and optical AGN samples in this study overlap only in a few objects.
This is because the OIII emission line cut selects 
radiatively efficient AGN while the radio-loud AGN population are mostly
radiatively inefficient. Therefore, the optical AGN sample presents radiatively
efficient radio-quiet AGN and the radio AGN sample presents radiatively
inefficient radio-loud AGN.

\section{Overall properties of the samples}
\label{sec:overal}

In this section, the general properties of the samples introduced in 
Section \ref{sec:Sample} are investigated. 

\subsection{Host galaxy properties}

The main properties of the galaxies in each sample
are displayed in Fig. 1. There are four samples of galaxies: 
i) galaxies in voids (blue), ii) isolated galaxies (green), iii) galaxies bounded in
groups including BGGs (pink) and iv) BGGs (red).
Galaxies in voids have statistically younger 
stellar populations (lower 4000$\AA$ break strengths) with more disk-like morphologies 
(lower concentrations). These parameters clearly show the late-type/passive 
bimodality in the population of galaxies caused by the galaxy evolution.
The vast majority of void galaxies are late-type. 
The BGG and group samples are dominated by passive galaxies. The population of 
isolated galaxies appears to have the same fraction of each type, 
shifted slightly toward the passive galaxies. 
Similarly, galaxies in voids are bluer and
have lower surface mass densities compared to the other three samples.
These results are consistent with those in the literature 
(Rojas et~al.\ 2004; Kreckel et~al.\ 2010; 
Ricciardelli et~al.\ 2017). In contrast,
the BGGs which are optically brighter by definition, are more massive with
higher black hole masses and ratios of black hole to stellar mass higher than those of the 
isolated and void galaxies.
The significance of the differences have been examined 
by the two-sided Kolmogorov-Smirnov (KS) test (Kolmogorov 1933). 
According to that, differences above the 3$\sigma$ confidence level satisfy
the condition
$$D_{m,n} > 1.818 \sqrt {\left(\frac{m+n}{m \times n} \right)},$$
where $D_{m,n}$ is the maximum difference between the cumulative distributions of the two samples.
Here, m and n are the sizes of the first and the second sample respectively. 
Only the differences above this level are reported in this paper.

\begin{figure*}
\centering
 \includegraphics[ scale=0.2]{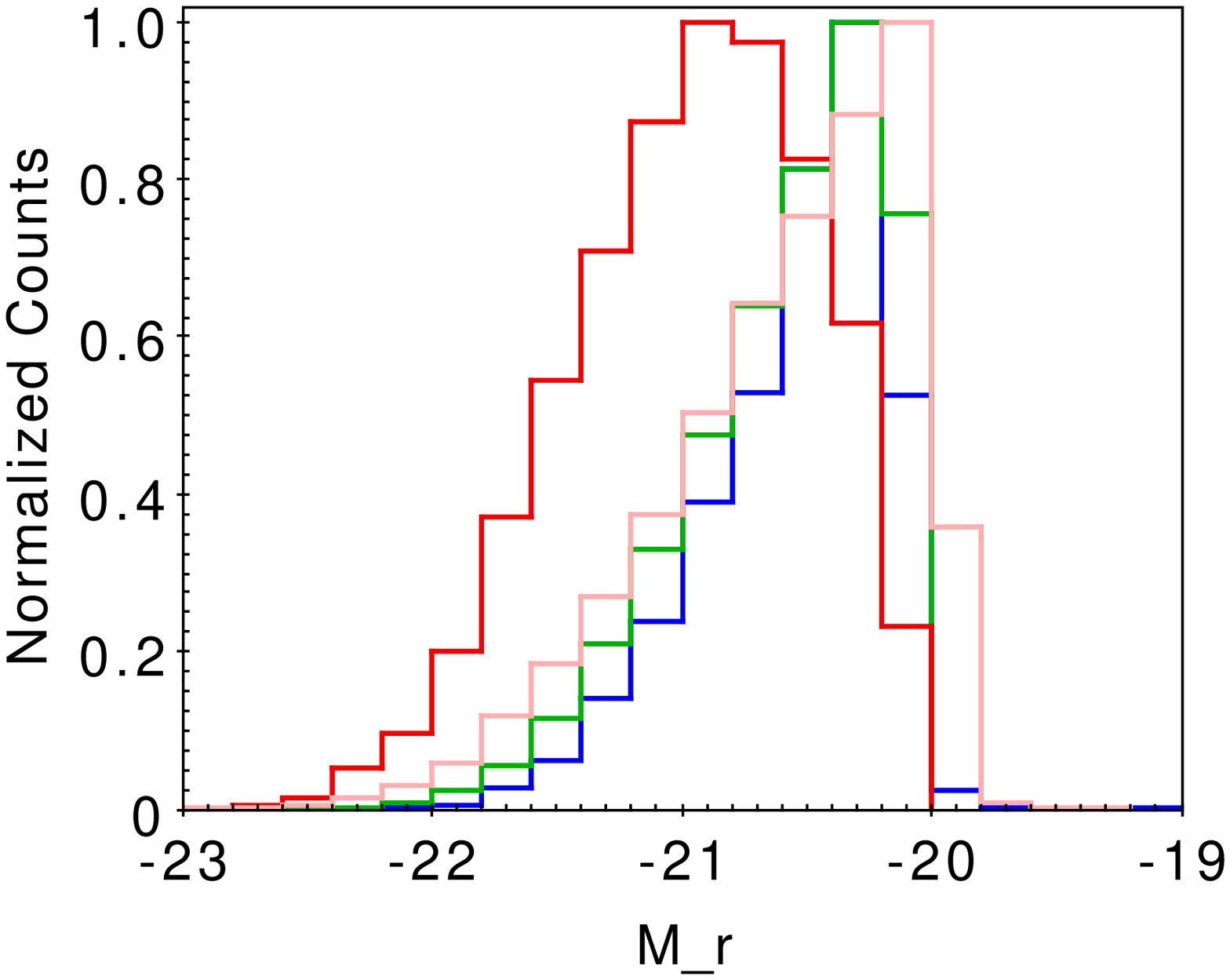}
 \includegraphics[ scale=0.2]{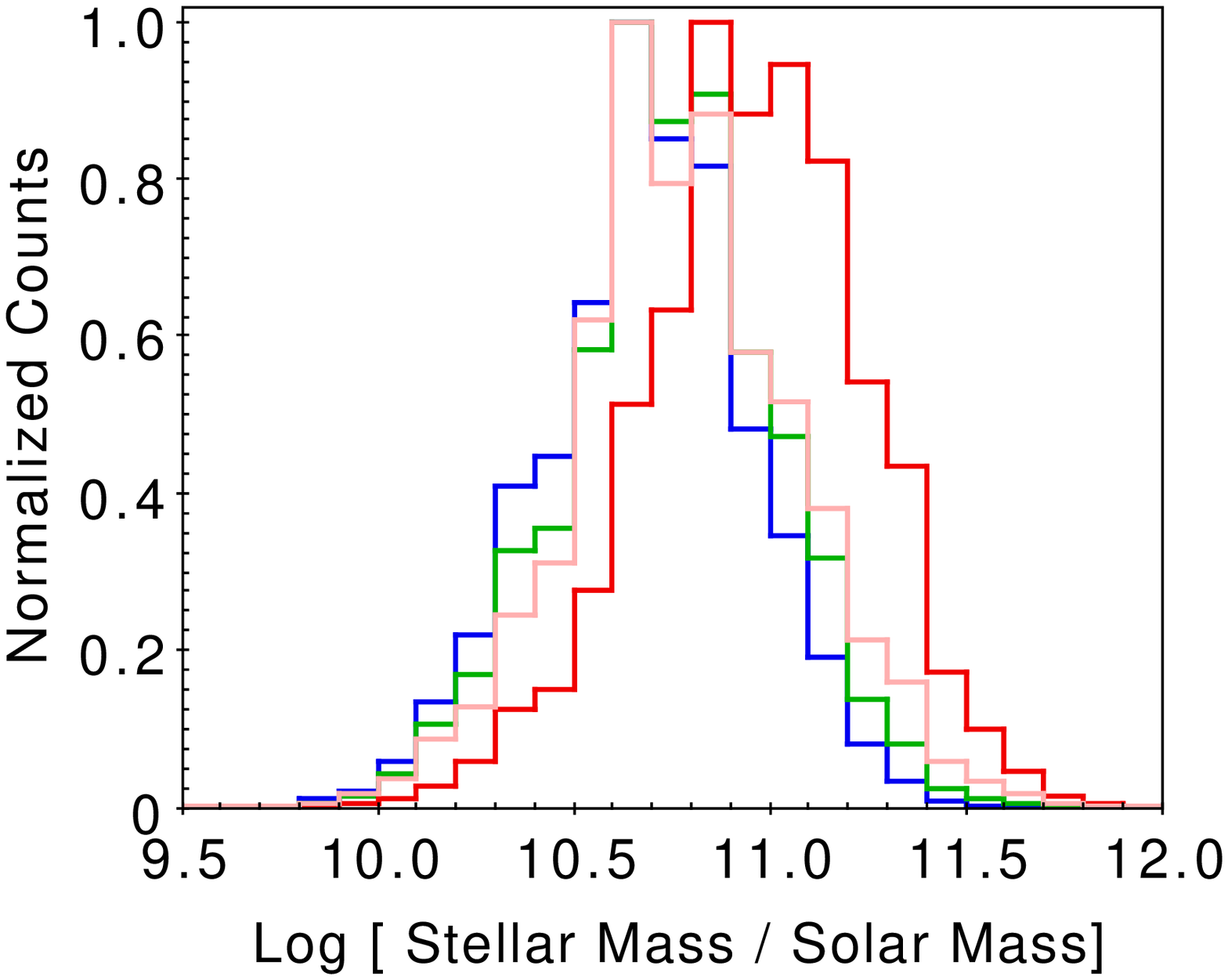}
 \includegraphics[ scale=0.2]{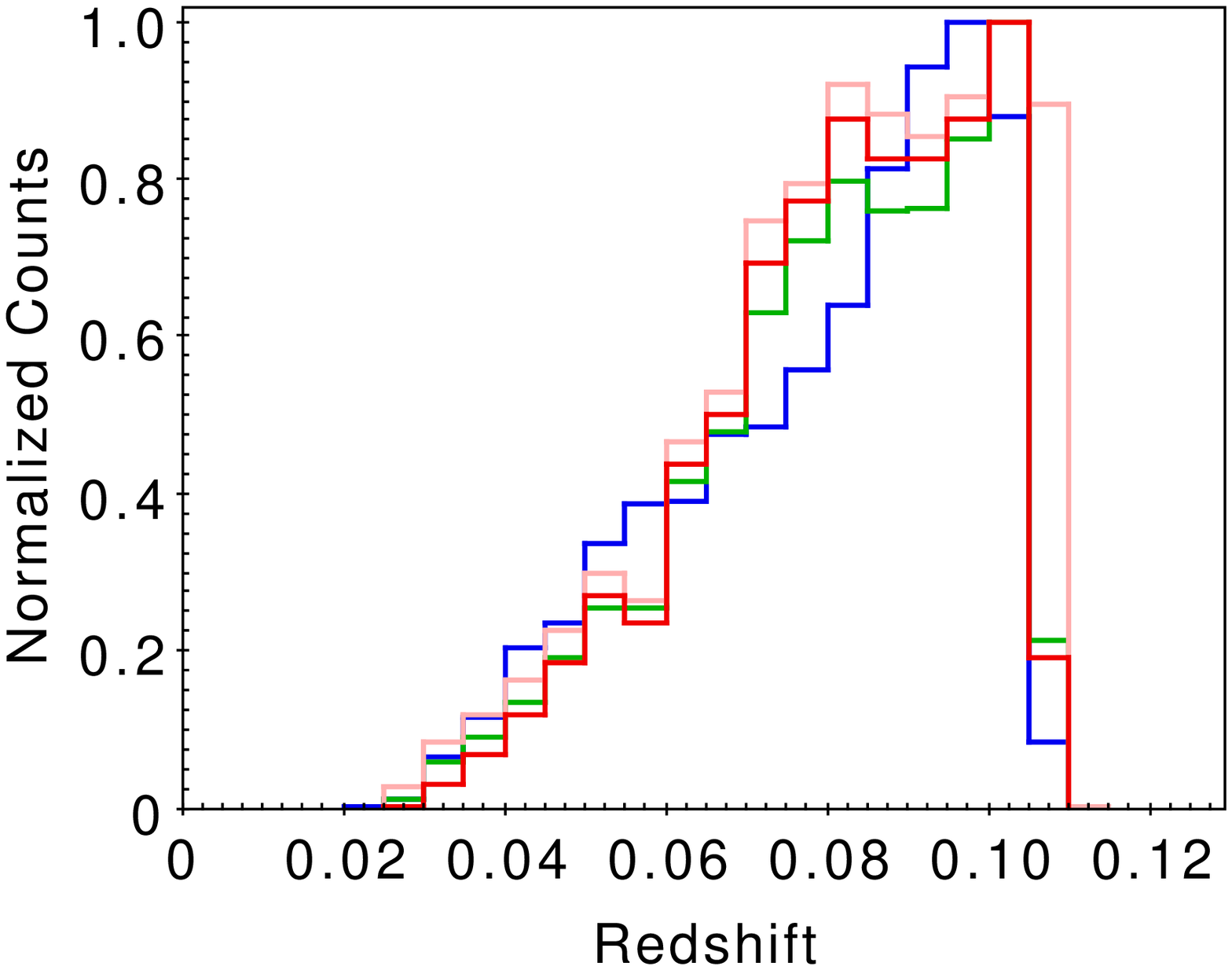}
 \includegraphics[ scale=0.2]{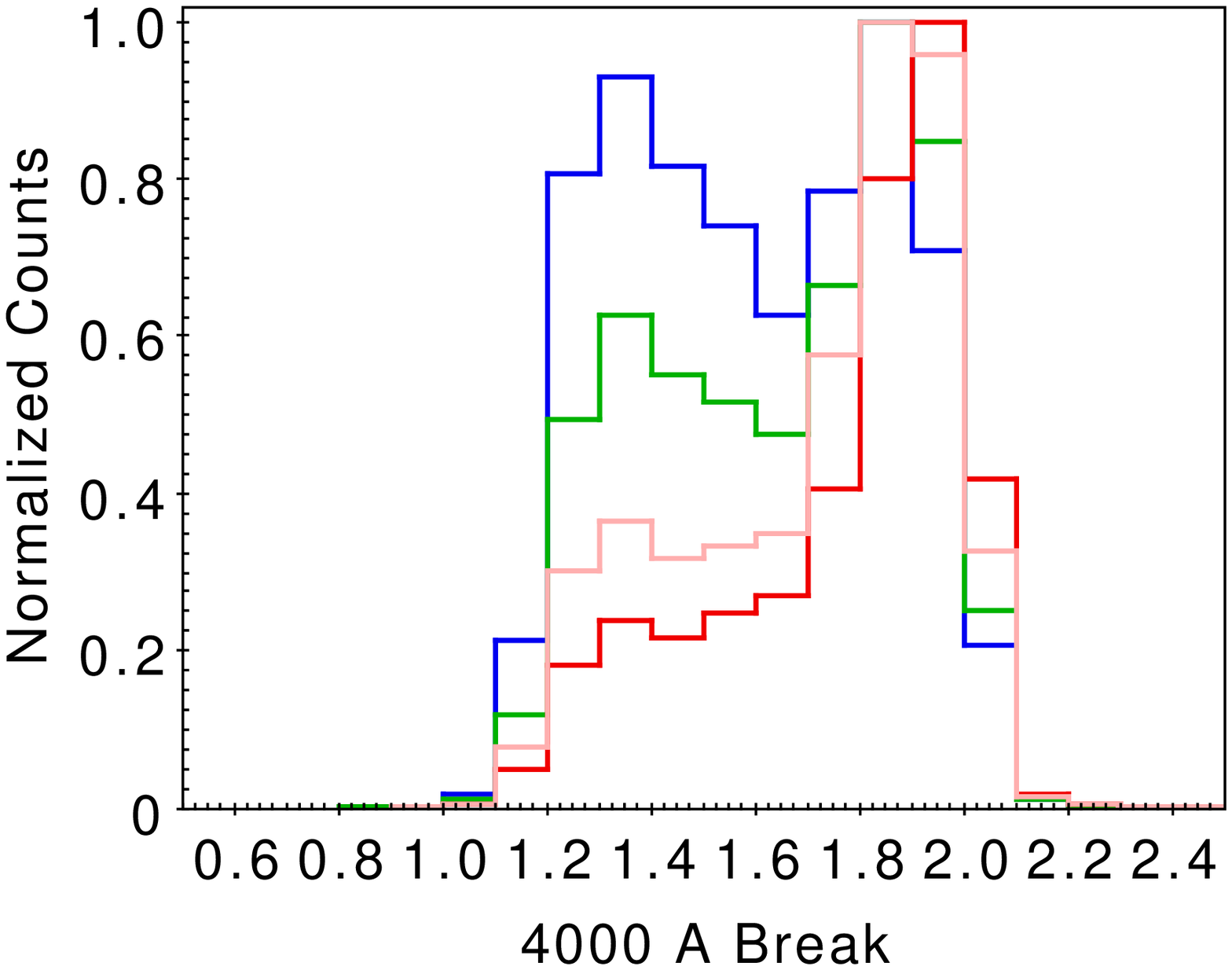}
 \includegraphics[ scale=0.2]{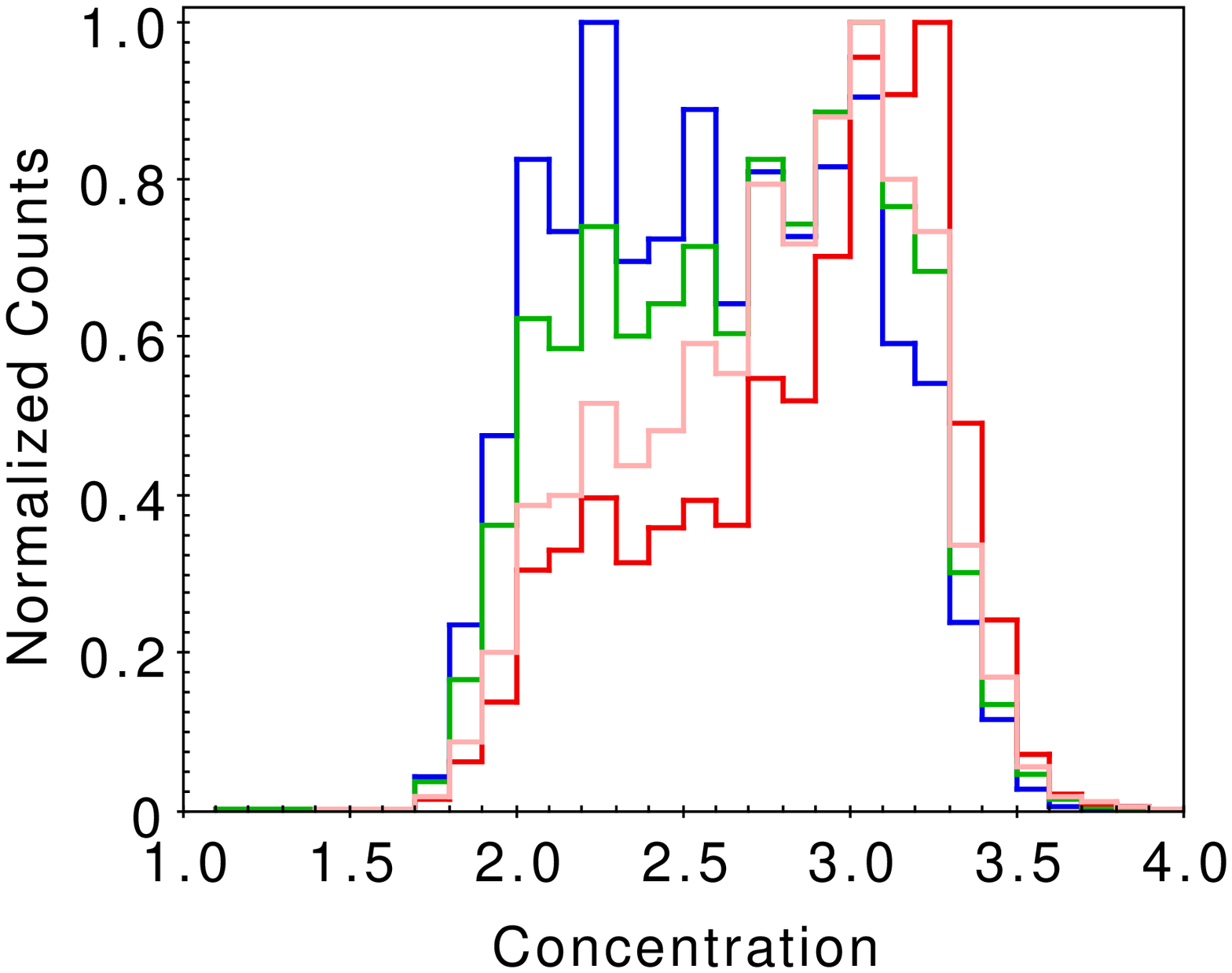}
 \includegraphics[ scale=0.2]{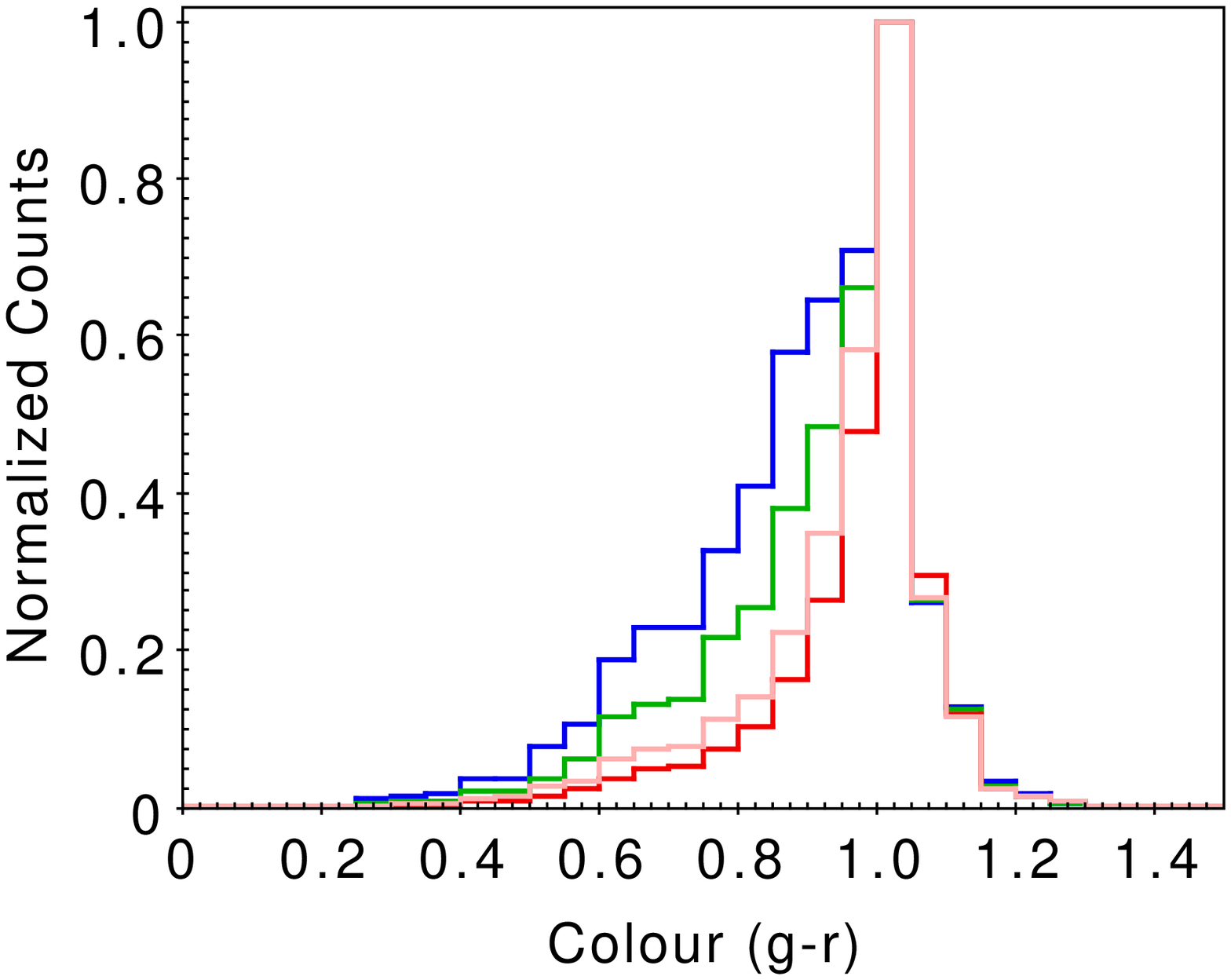}
 \includegraphics[ scale=0.2]{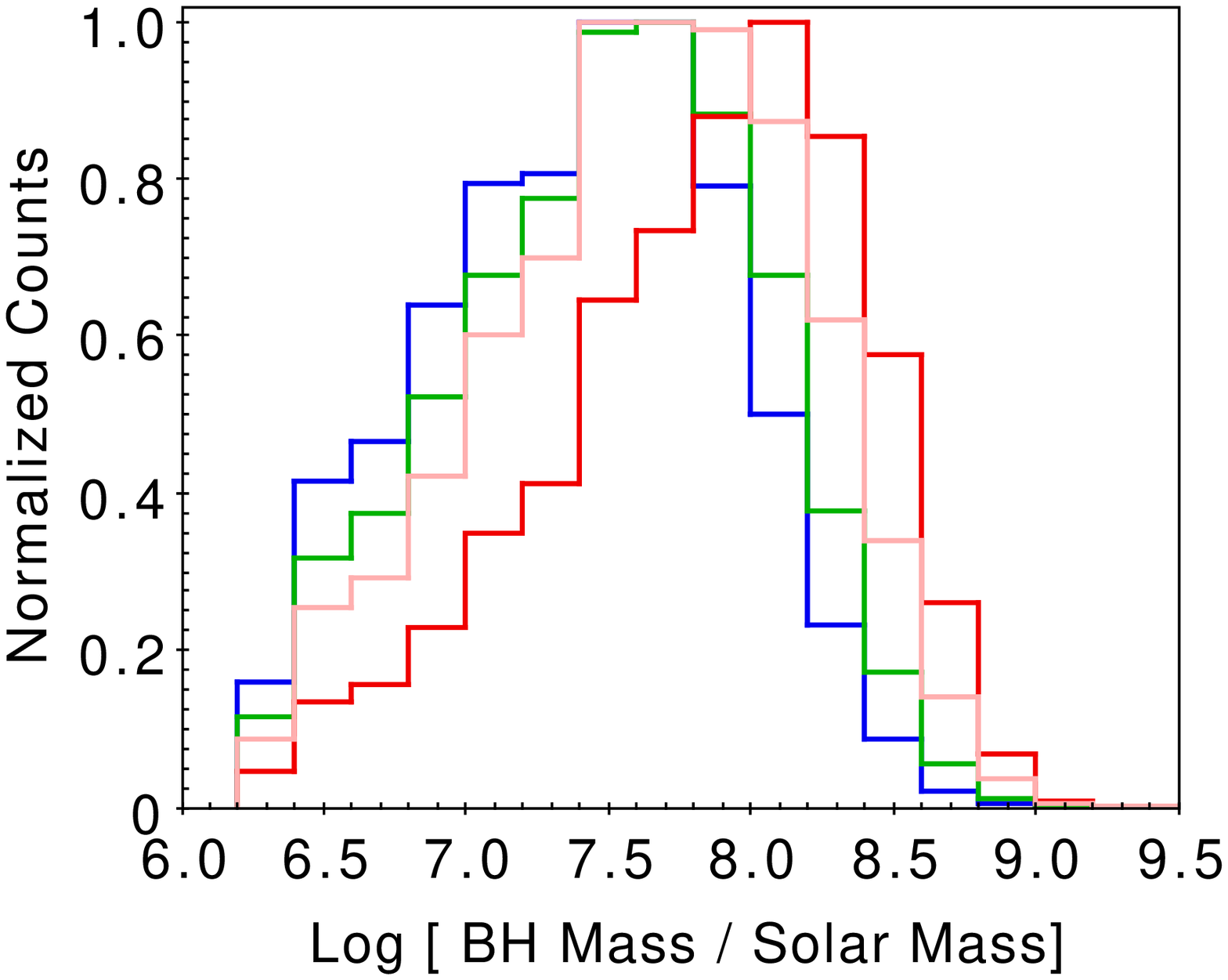}
 \includegraphics[ scale=0.2]{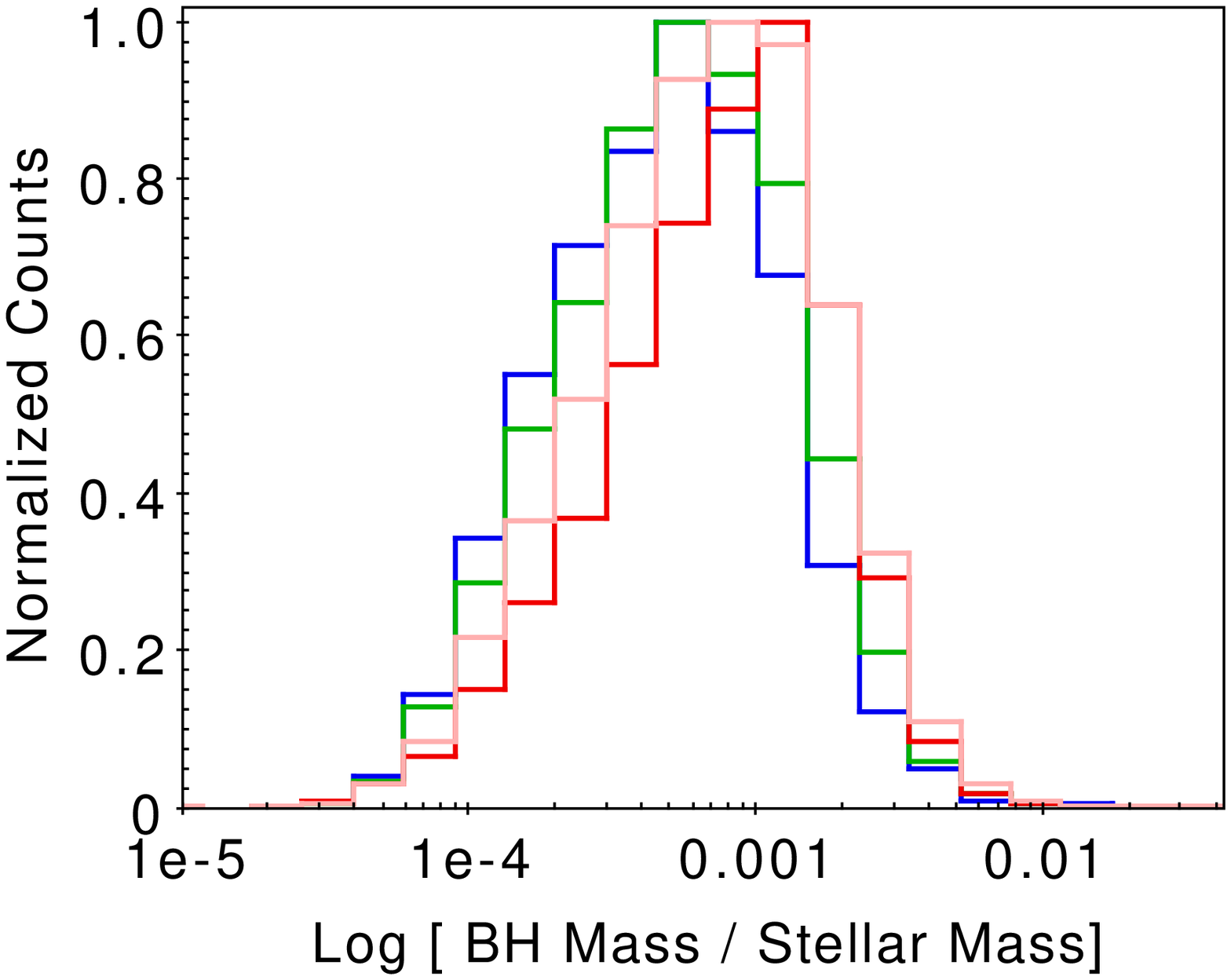}
 \includegraphics[ scale=0.2]{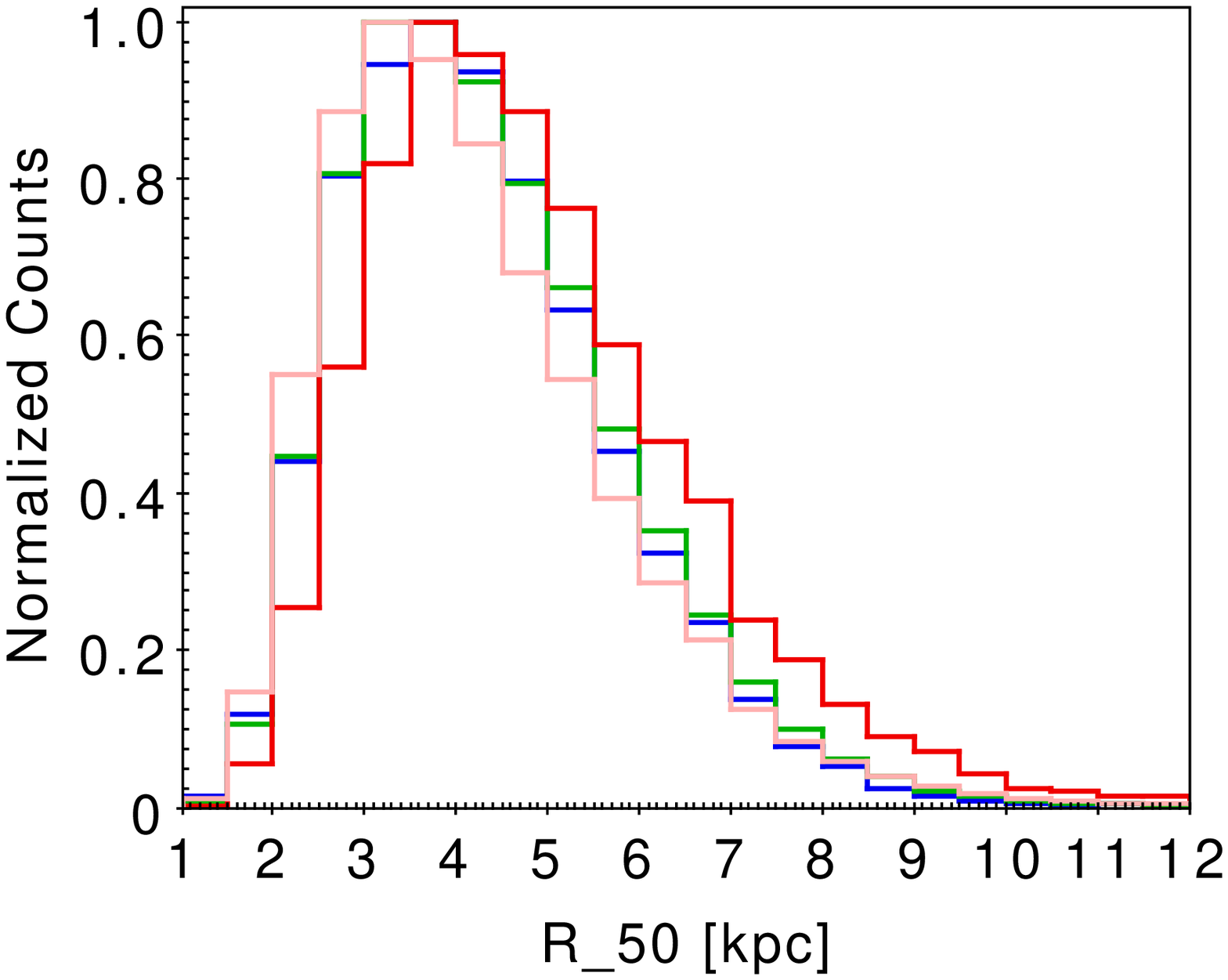}
 \includegraphics[ scale=0.2]{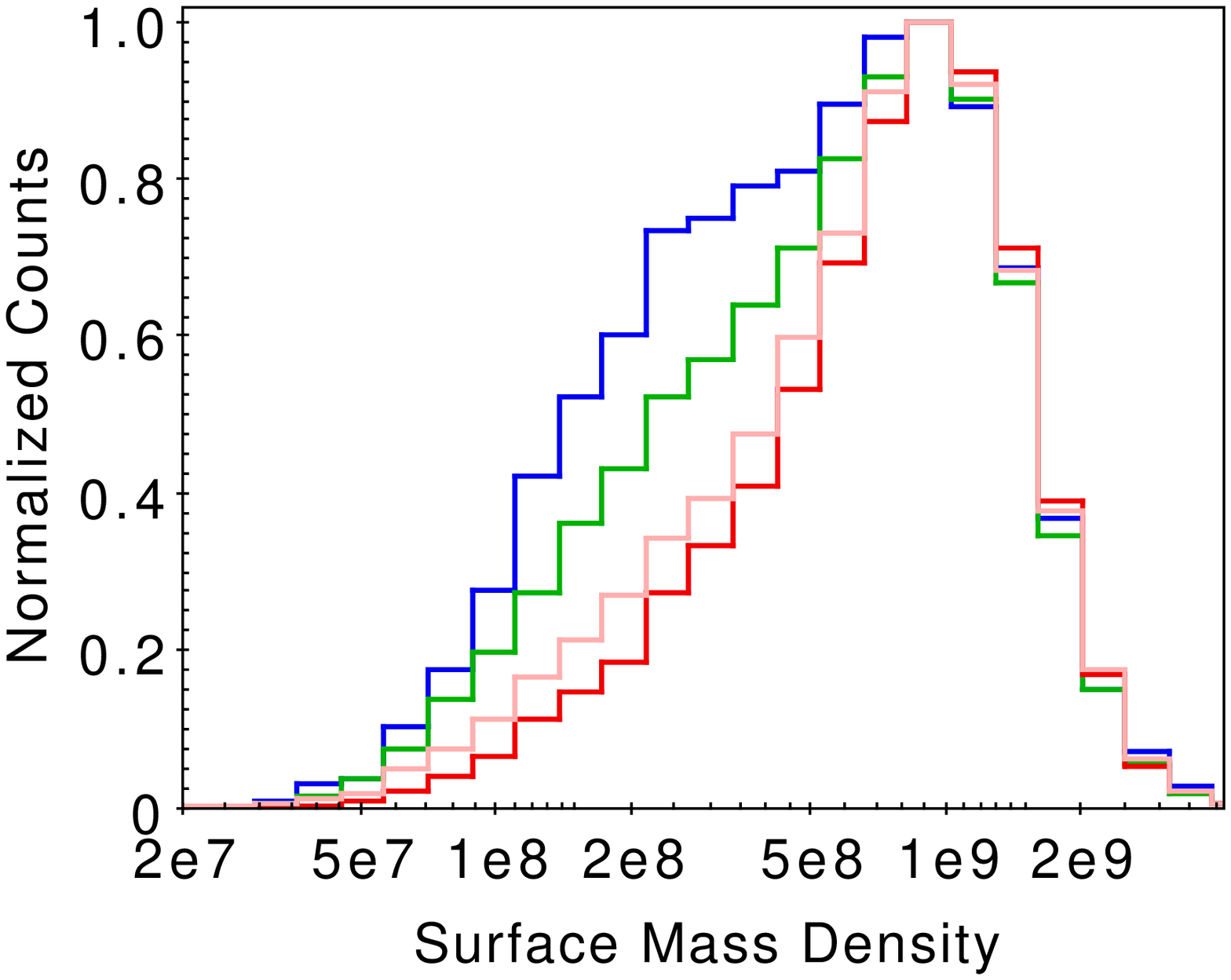}
\figcaption{Histograms of the optical properties of galaxies. The colours represent 
galaxies in voids (blue), isolated galaxies (green),
galaxies in groups (pink) and BGGs (red).
\label{figophis}}
 \end{figure*}

\begin{figure*}
\centering
 \includegraphics[ scale=0.2]{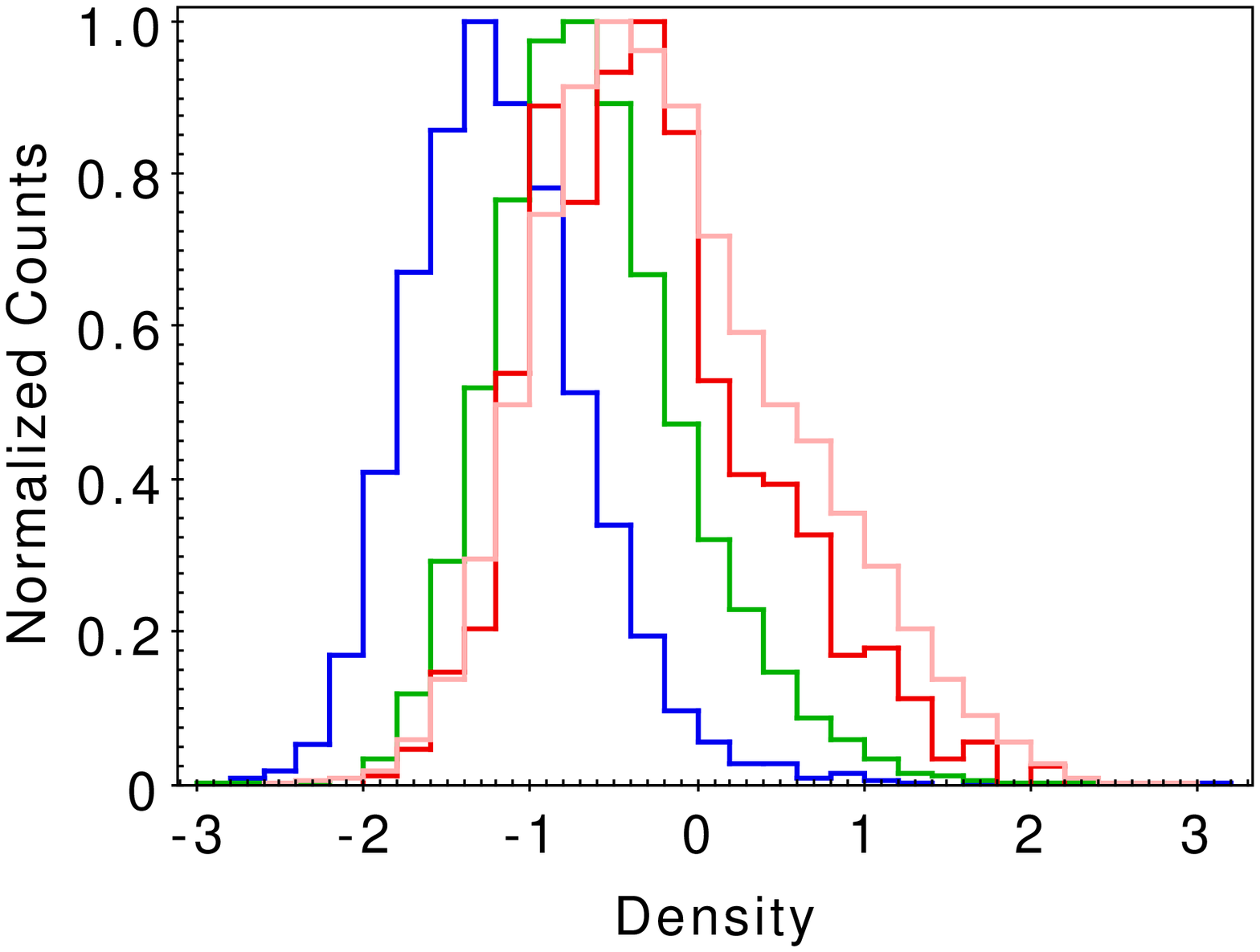}
 \includegraphics[ scale=0.2]{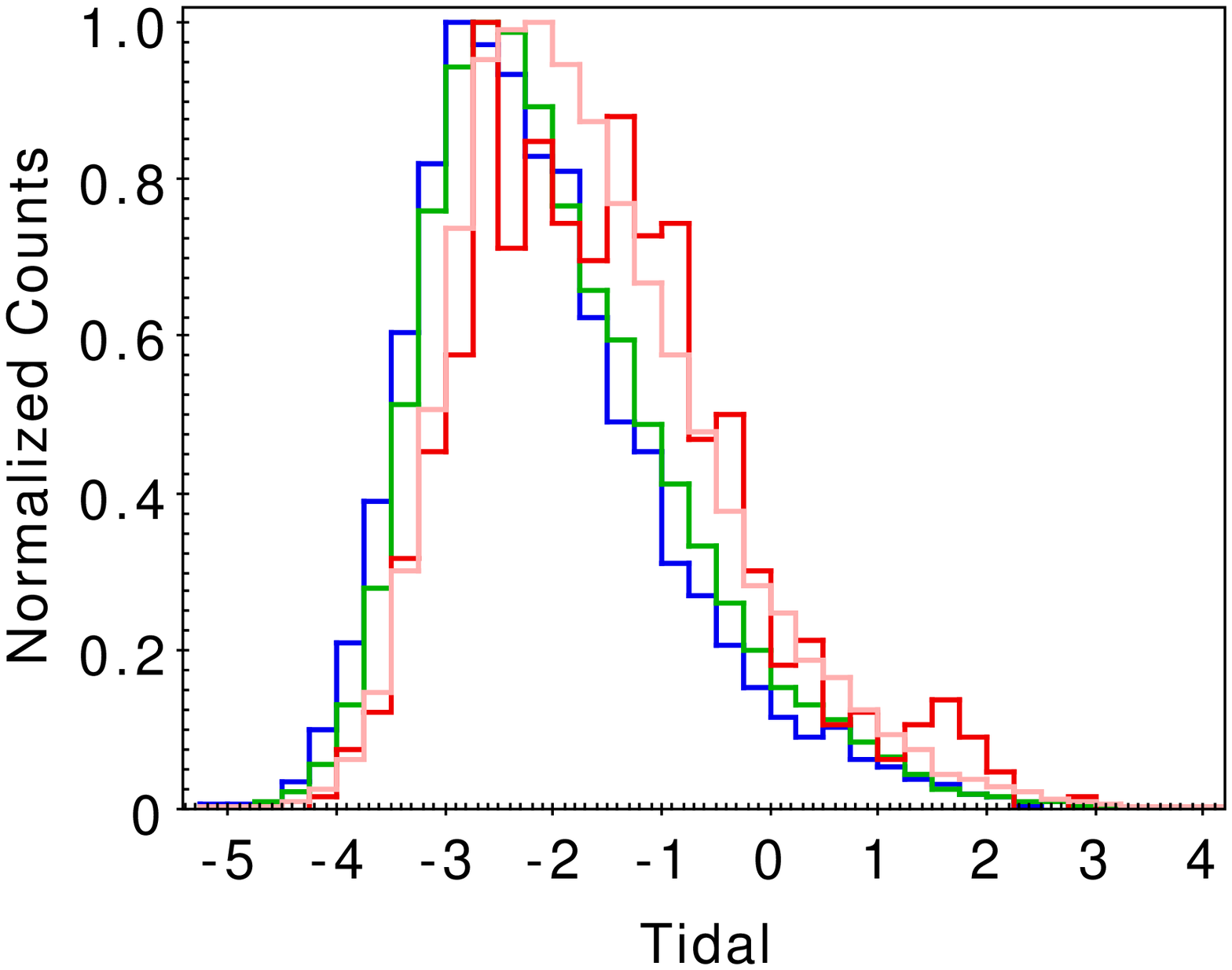}
 \includegraphics[ scale=0.2]{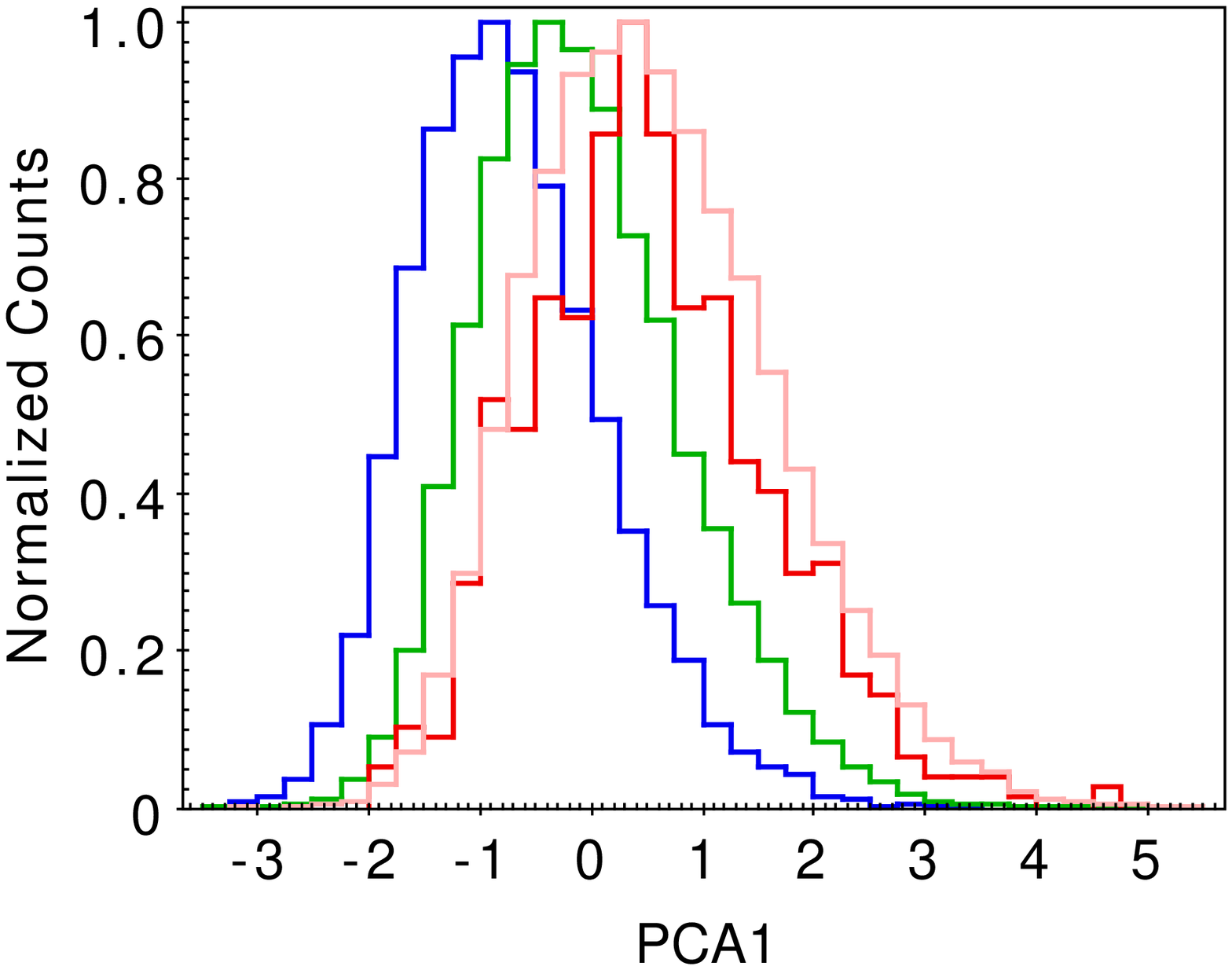}
 \includegraphics[ scale=0.2]{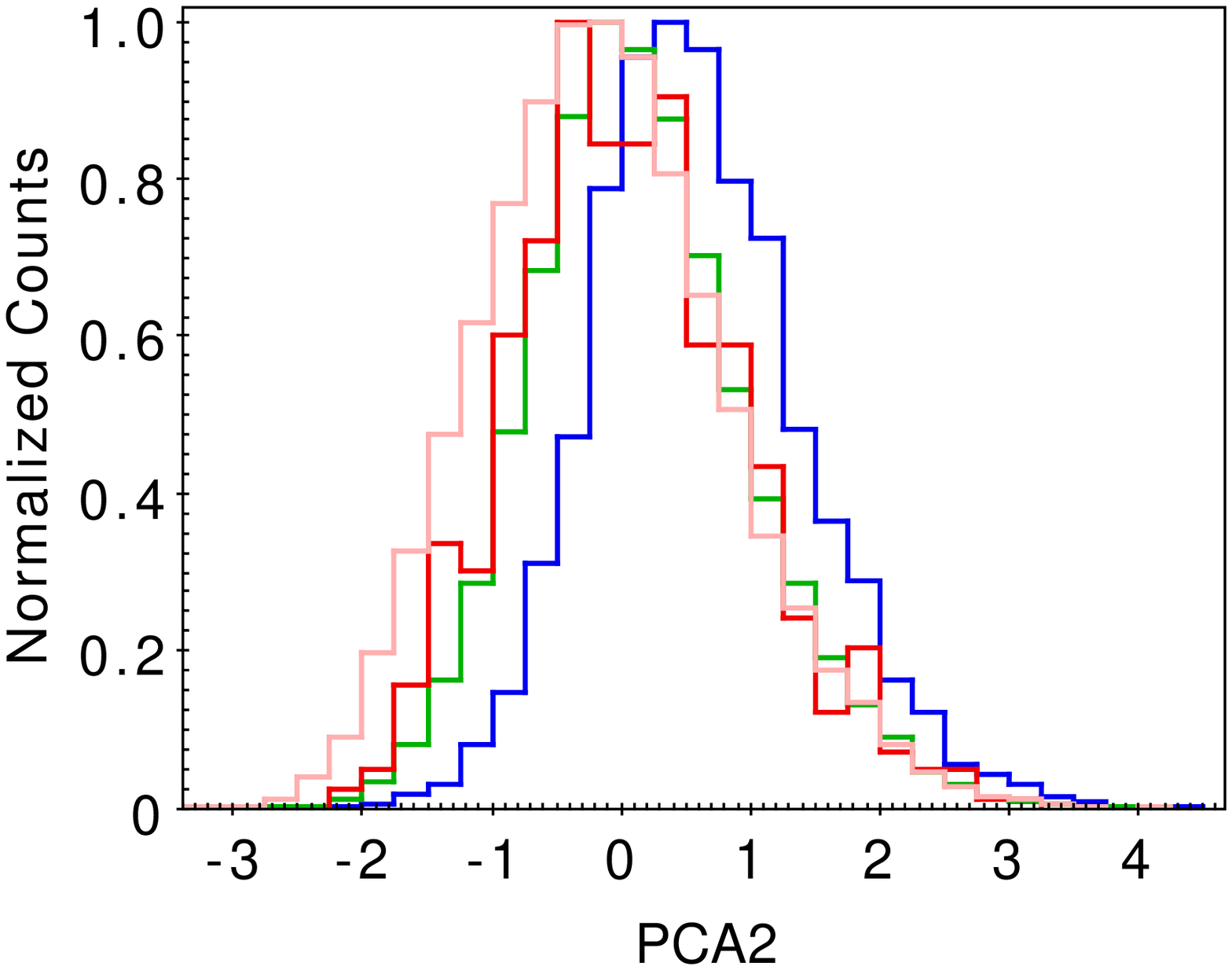}
 \includegraphics[ scale=0.2]{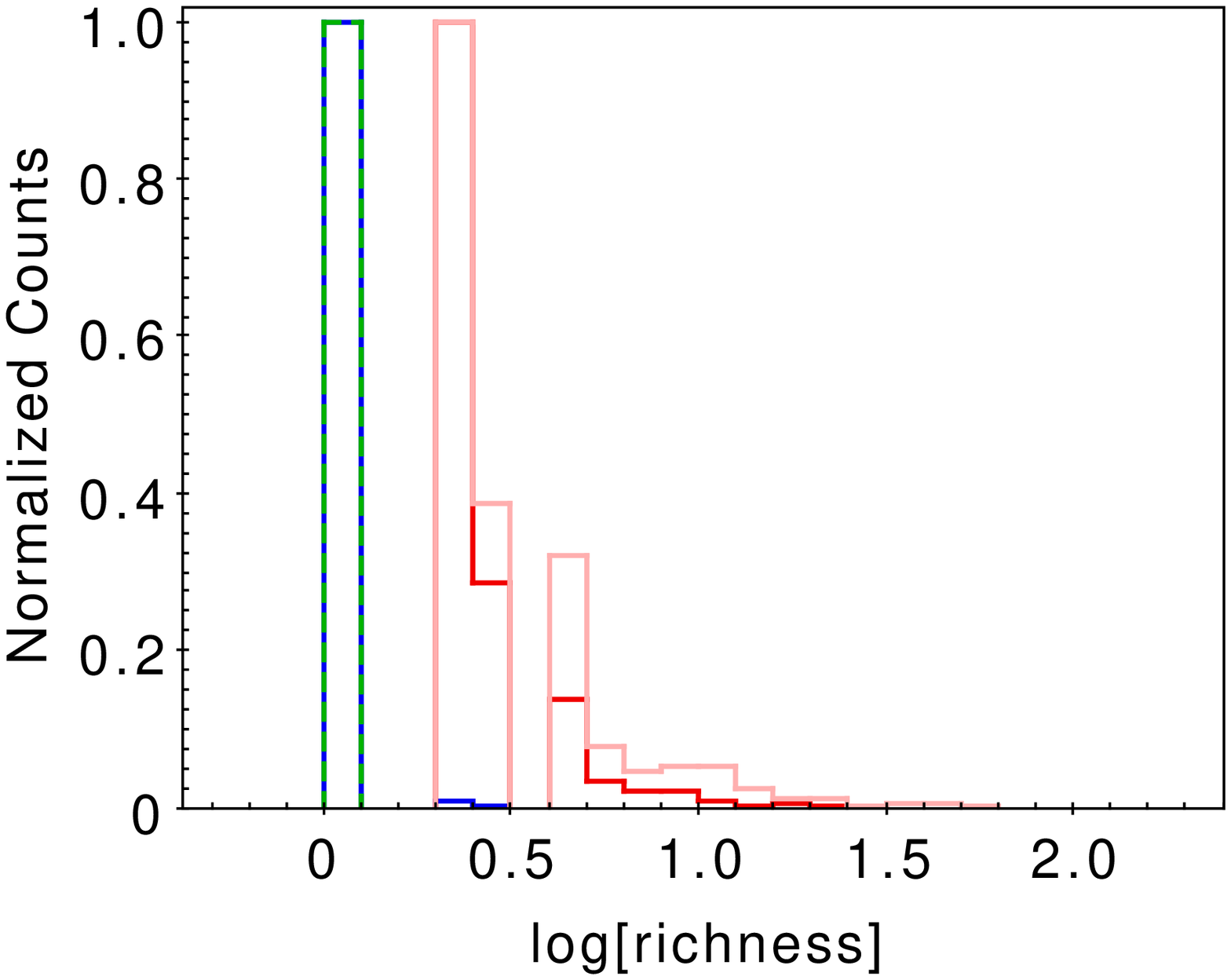}
 \includegraphics[ scale=0.2]{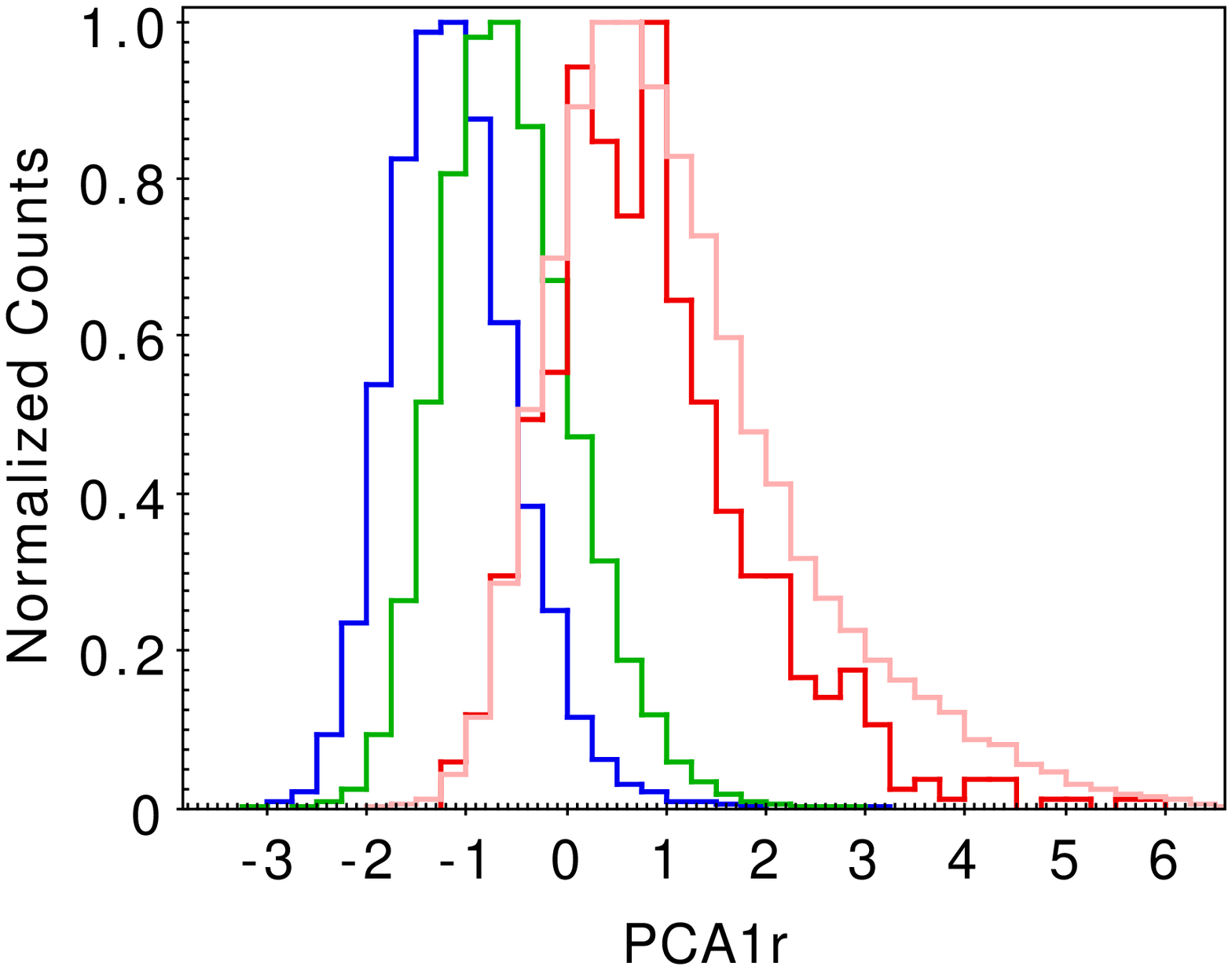}
 \includegraphics[ scale=0.2]{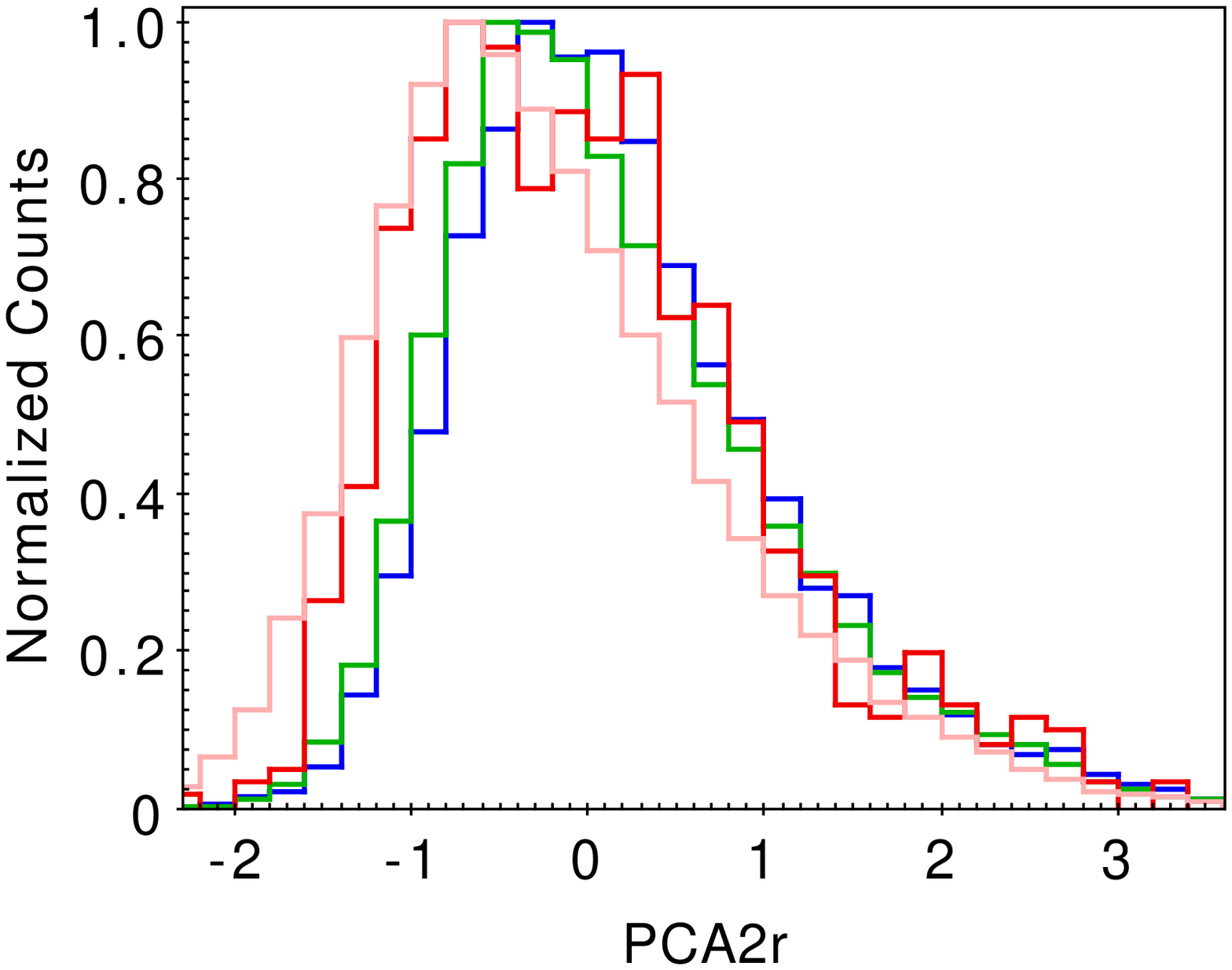}
 \includegraphics[ scale=0.2]{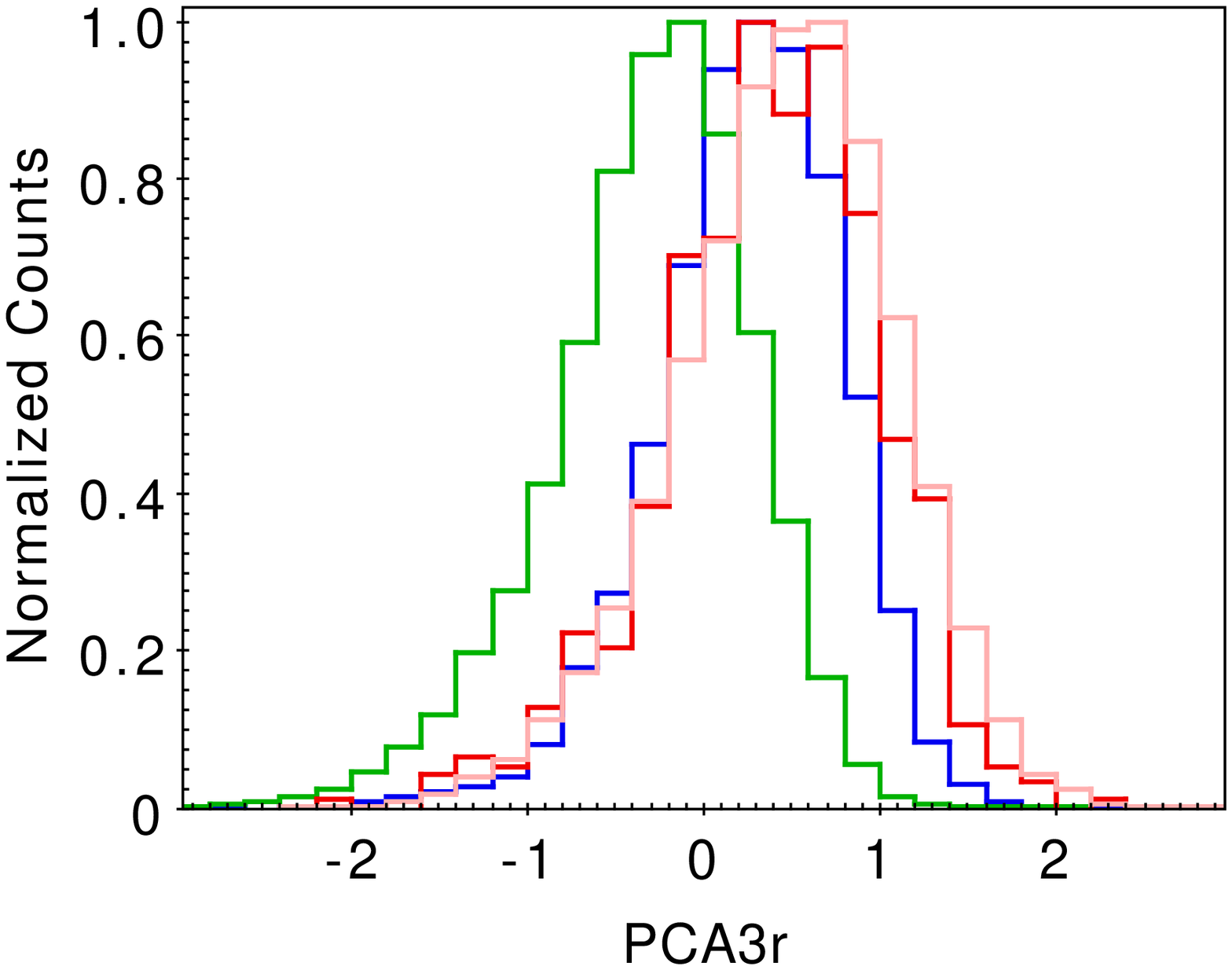}
\figcaption{Histograms of the environmental properties of galaxies. The colours represent different samples
as in Fig. \ref{figophis}.
\label{figopen}}
 \end{figure*}

\subsection{Environmental properties}

\begin{figure*}
\centering
 \includegraphics[scale=0.35]{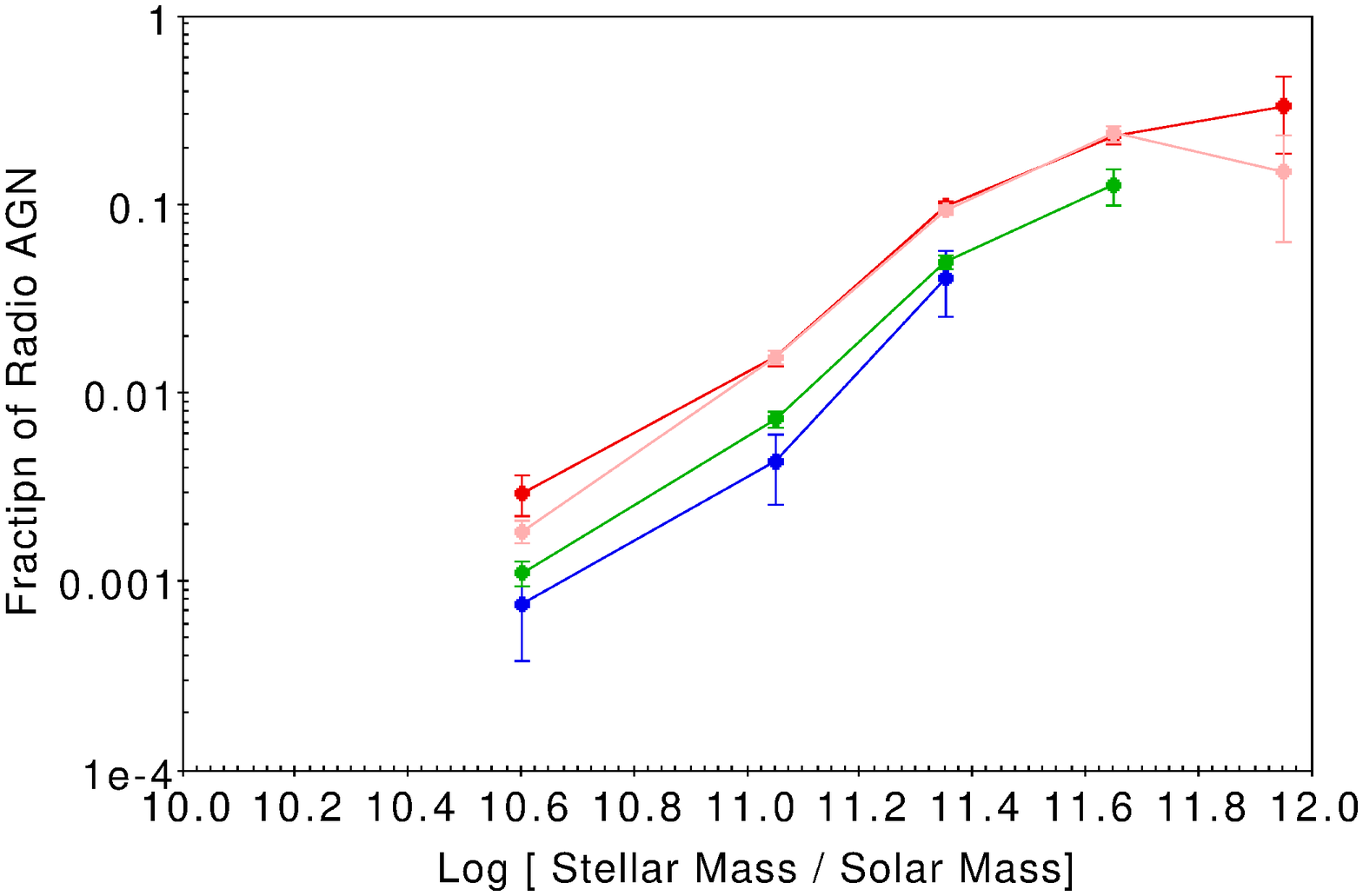}
 \includegraphics[scale=0.35]{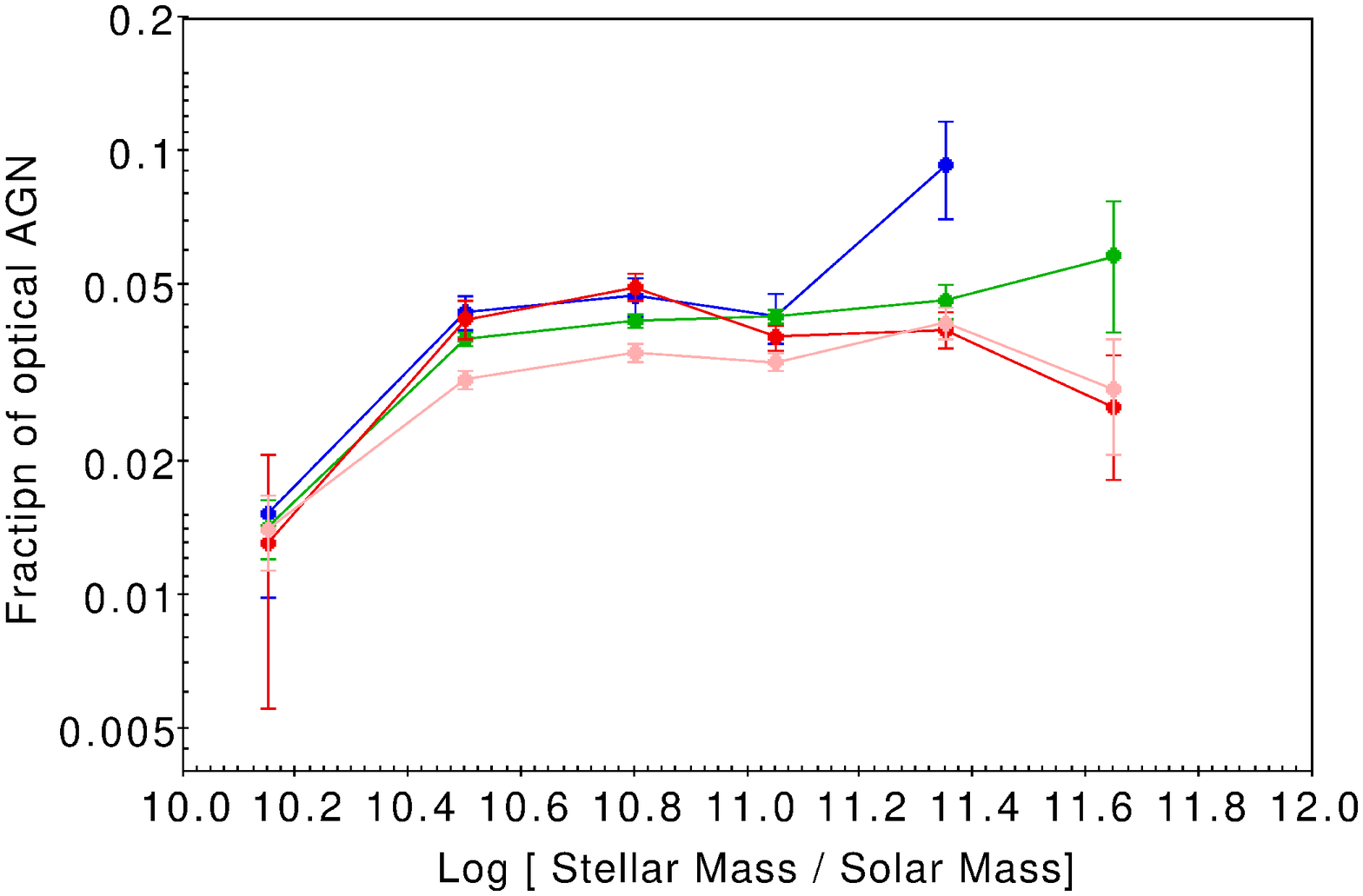}
\figcaption{Fraction of galaxies 
hosting radio (left-hand panel) and optical (right-hand panel)
AGN in each stellar mass bin and for different environments.  
 Colours represent four environmentally different samples as in Fig. \ref{figophis}.
Results are biased by the host galaxy type.
\label{figen}}
 \end{figure*}

To quantify the environmental properties of each sample, 
the distributions of density, 
tidal interaction and the PCA parameters are displayed in Fig. 2. This allows us to 
find the differences between the galaxy samples specifically 
comparing the environment of the isolated galaxies with the void galaxies 
and the group galaxies with the BGGs.
The density distributions show that the void galaxies are in the most
underdense regions, then the isolated galaxies and finally 
the group galaxies or BGGs have the highest-density environments. 
In terms of tidal interaction, small differences are found between 
different samples. Group galaxies and the BGGs experience higher tidal forces
than isolated or void galaxies. PCA1 is the main PCA component that is in the direction of 
density-tidal correlation and 
follows the same behaviour as the density does. Therefore, the overall 
interaction level for the galaxies is the highest in the BGGs and group member galaxies and
is the lowest in void galaxies. This is also confirmed by PCA1r. 

PCA2 which has
negative contribution of density and positive contribution
of tidal interaction includes information in the direction perpendicular to the PCA1.
According to that, galaxies in voids have the highest PCA2 values. Therefore, 
despite their surrounding underdense environment, void galaxies 
 still experience tidal forces from one-on-one interactions probably from their local
environments. This might be caused by having few nearby galaxies. It also  
confirms the results in Kreckel et~al.\ (2010) who show void galaxies are gas-rich
and display the evidence of gas interactions with the environment.
The PCA3r parameter further confirms this. Void galaxies along with galaxies in group and BGGs
have higher PCA3r than isolated galaxies. A reason for that is a population of poor 
galaxy groups in voids. The samples show a small difference in PCA2r. 

In summary,  
density, PCA1 and PCA1r as indicators for overall interaction 
level show galaxies in voids to be the lowest interacting galaxies. PCA2 and PCA3r which
are more sensitive to local interactions show that galaxies in void 
are still significantly influenced by their nearby galaxies.   
Isolated and void galaxies display different environmental properties. 
Isolated galaxies are located in higher density regions 
but experience lower levels of local interaction than void galaxies.
There is no significant difference, on-average, between group galaxy members and the BGG
when environmental properties have been accounted for.

\subsection{AGN activity: A general view}

The fractions of galaxies hosting radio (left-hand panel) and optical (right-hand panel)
AGN in different environments with respect to the stellar mass are shown in Fig. \ref{figen}.
The colours are defined as in Fig. \ref{figophis}. 
The fractions are n/N and the error bars are calculated 
using a Poissonian approach as  $\sqrt{n}$/N where
 n and N are the number of radio/optical AGN and the total number 
of galaxies in each stellar mass bin respectively.
A larger stellar mass bin width has been used to increase the signal to noise ratio (S/N)
for the radio AGN fraction.
The plots show the general trend of AGN activity 
with respect to the stellar mass and environment before
removing the biases caused by the galaxy type and the results 
can be compared with some of the previous ones in the literature.
The fraction of the radio-mode AGN increases
significantly with increasing stellar mass. A recent study shows that 
this fraction reaches as high as hundred percent for very massive galaxies (Sabater et~al.\ 2019). 
Galaxies in voids show the lowest AGN fraction. 
Isolated galaxies have slightly higher radio AGN fraction than void galaxies. Galaxies
in groups and the BGGs show the highest level of radio-mode AGN activity. There is 
only a small difference between the result for the BGGs and the group member galaxies.

The optical AGN activity does
not show strong dependence on the stellar mass when different 
environments are considered. This is investigated further in Sections 
\ref{sec:colour} and \ref{sec:R}. The fraction of optical AGN is higher in void galaxies than 
in isolated and group galaxies. Isolated galaxies have higher fractions of
optical AGN activity than galaxies in groups. The same result has been reported by
Kauffmann et~al.\ 2004 who detected lower 
numbers of optical AGN in higher-density environments. 
There is a noticeable fraction of optical AGN for the low-mass
BGGs compared to the group galaxies. The optical AGN fraction in BGGs decreases toward
higher stellar masses such that the AGN fraction for massive BGGs are lower than the AGN fraction in 
voids and isolated galaxies consistent with the results in von der Linden et~al.\ (2007). 
The results of this section are investigated further
in Section \ref{sec:colour} to find out if they are biased by the galaxy 
type. Each sample has different fraction of blue, green and red galaxies.
A higher fraction of blue star forming galaxies in voids and isolated galaxies than in group galaxies 
(as shown in Fig. 1) leads to a higher fraction of optical AGN.

\section {AGN activity: dependence on the stellar mass and colour}
\label{sec:colour}

\begin{figure*}
\centering
 \includegraphics[scale=0.35]{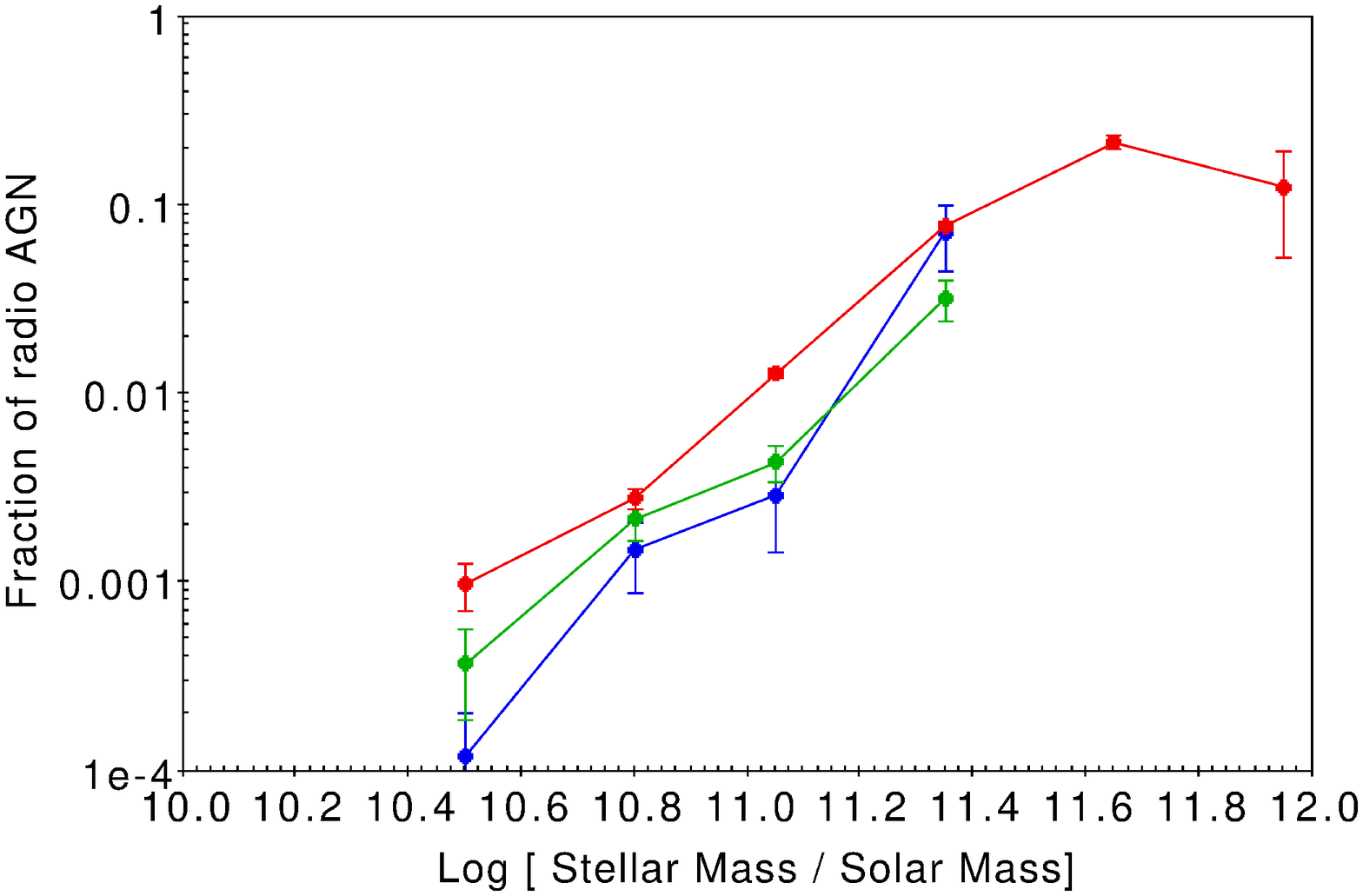}
 \includegraphics[scale=0.35]{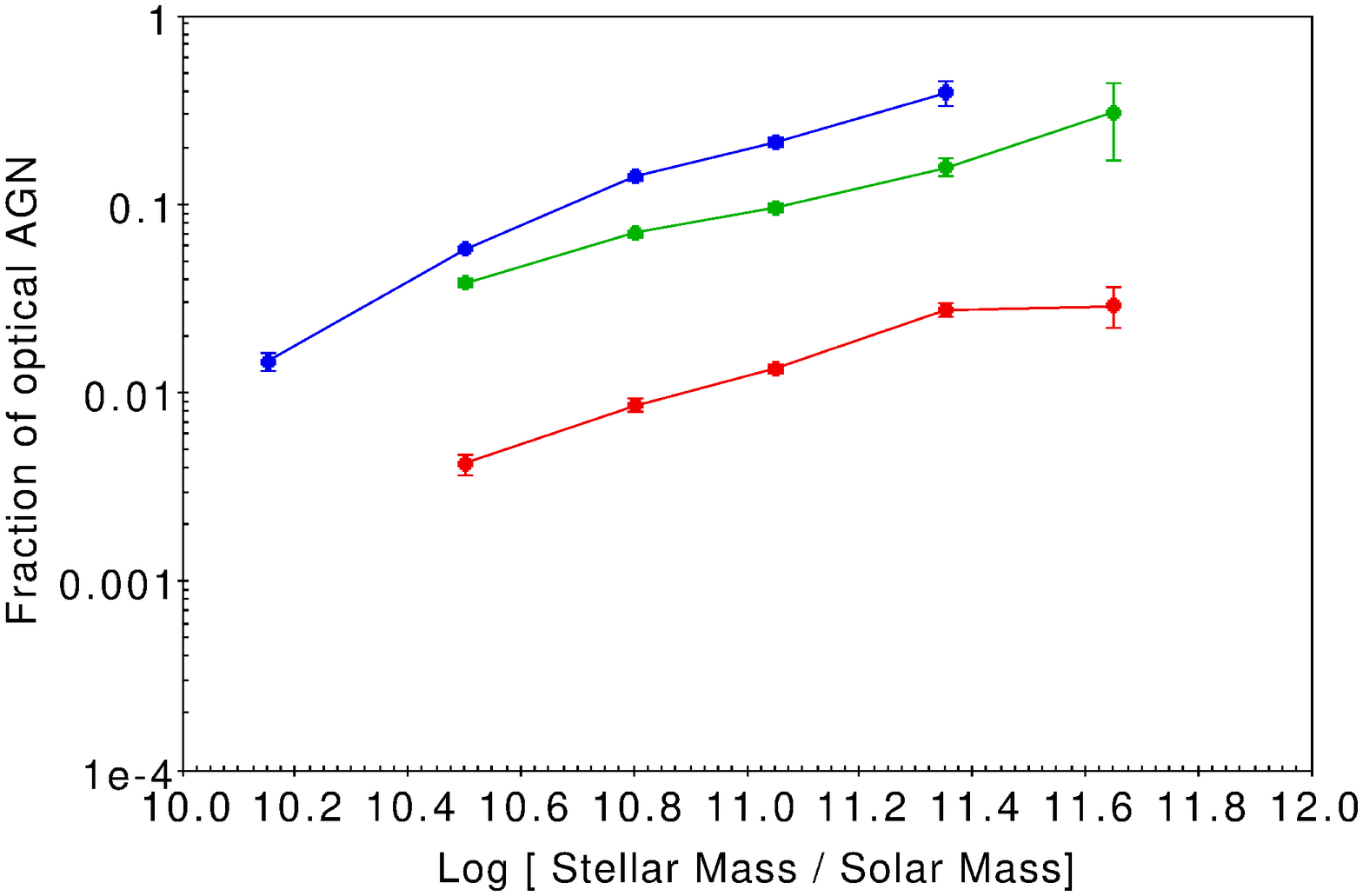}
\figcaption{Fraction of blue (blue line), green (green line) and red (red line) galaxies 
hosting radio (left-hand panel) and optical (right-hand panel)
AGN in each stellar mass bin. Galaxy colour is defined in Sec. \ref{sec:colour}. 
\label{figop}}
 \end{figure*}

In this section, the relation between AGN activity and the host galaxy properties
is studied. It is necessary to separate the impact of these properties 
from the environmental properties on AGN activity in order to ensure that
any dependence of AGN activity on the environment is not biased by 
them. One major set of properties, displayed in Fig. 1, includes
r-band magnitudes, stellar masses, black hole masses and galaxies optical sizes (R$_{50}$).
This set indicates how large galaxies are.
The rest of the properties which include 4000$\AA$ break strengths,
concentrations, g-r colours, ratio of black hole to stellar mass and surface mass densities
represent galaxies types. The stellar mass and 4000$\AA$ break strength have been selected from each
set of properties to be used to construct similar host galaxy samples.
The stellar mass has been selected since it is the main property 
of galaxies and most of the galaxies properties are functions of the stellar mass.
4000$\AA$ break strength has been selected since it
represents galaxy type bimodality with a wider separation than other 
properties such as g-r colour. The bimodality in the  
distribution of 4000$\AA$ break strengths of galaxies is shown in Fig 1.
The blue, green and red galaxies are defined based on the
4000$\AA$ break strength discussed in Janssen et~al.\ (2012) as follows:   
\begin{description}
\item  \space \space  blue galaxies:  \space \space \space \space \space \space \space \space \space 4000$\AA$ break $\leq$ 1.45
\item   \space \space  green galaxies: \space 1.45 $\leq$ 4000$\AA$ break $\leq$ 1.7 
\item    \space \space  red galaxies:   \space  \space \space \space  1.7 $\leq$ 4000$\AA$ break 
\end{description}

In the following, the dependence of AGN activity on the 
stellar mass and galaxy colour, as defined above, will be shown
regardless of the galaxies environments. This helps us to find out how each mode of AGN activity
changes with either the stellar mass or colour. The potential biases that may be caused by
the rest of the properties will then be explored in a fixed stellar mass-colour bin. 

\begin{figure*}
\centering
 \includegraphics[ scale=0.35]{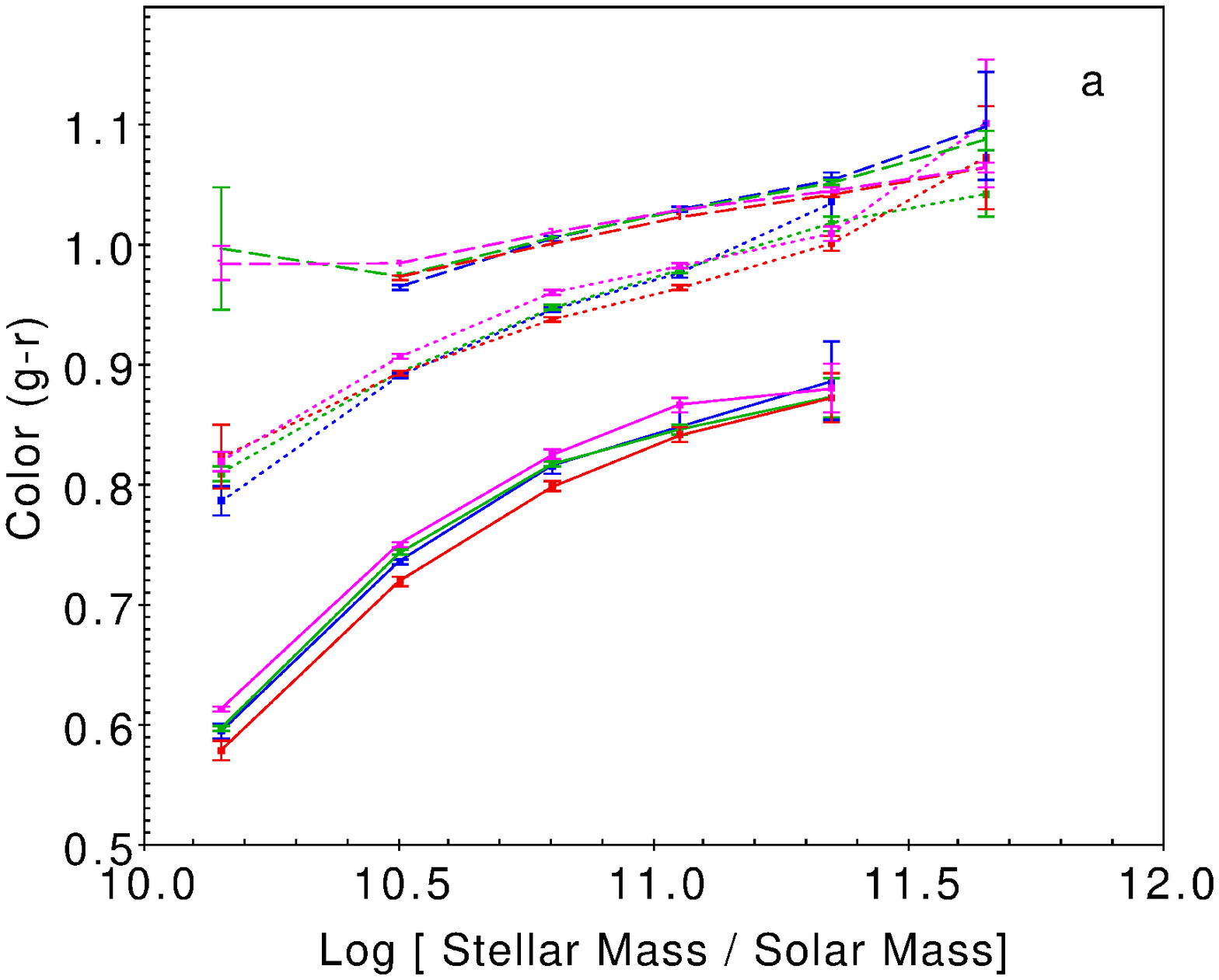}
 \includegraphics[ scale=0.35]{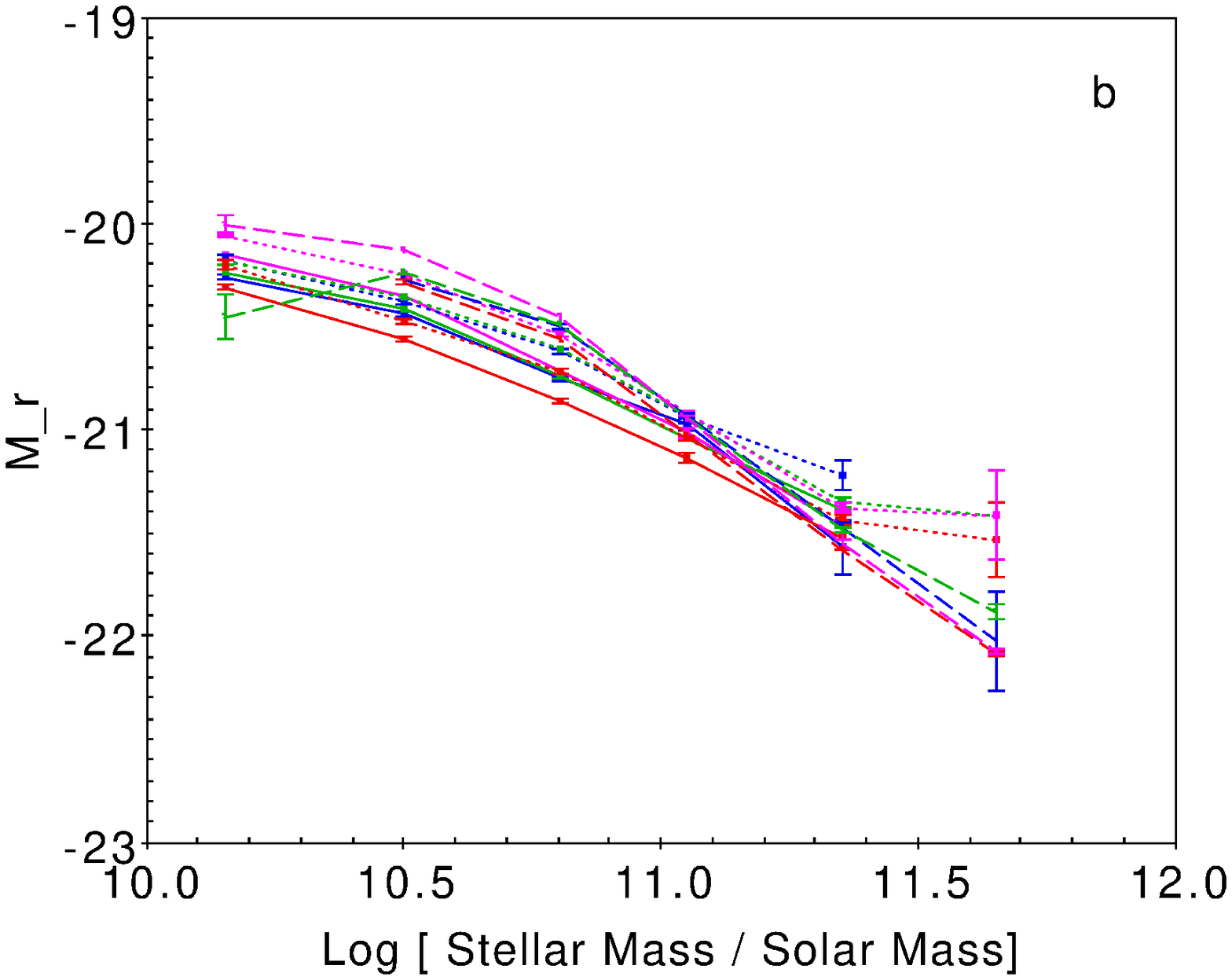}
 \includegraphics[ scale=0.35]{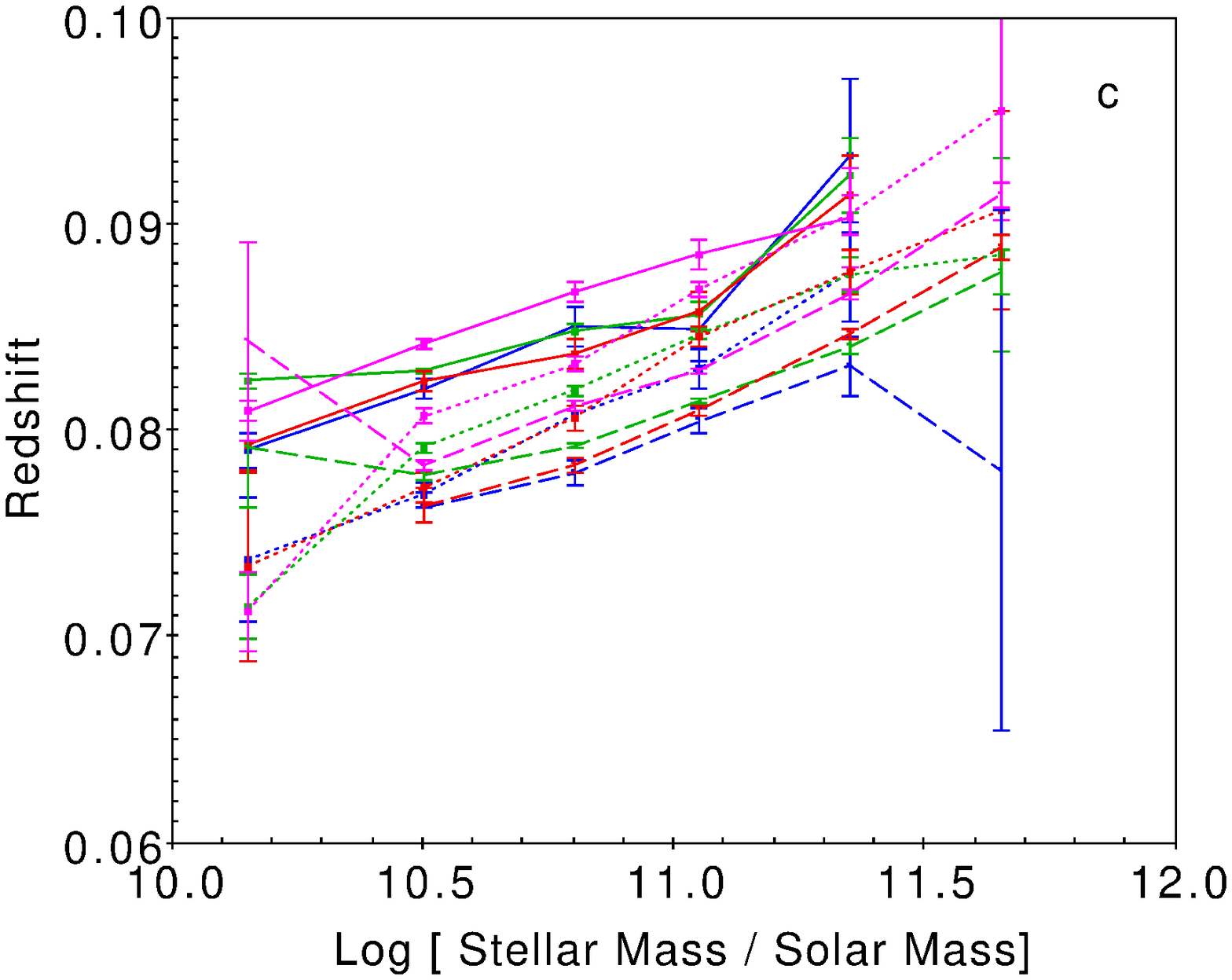}
 \includegraphics[ scale=0.35]{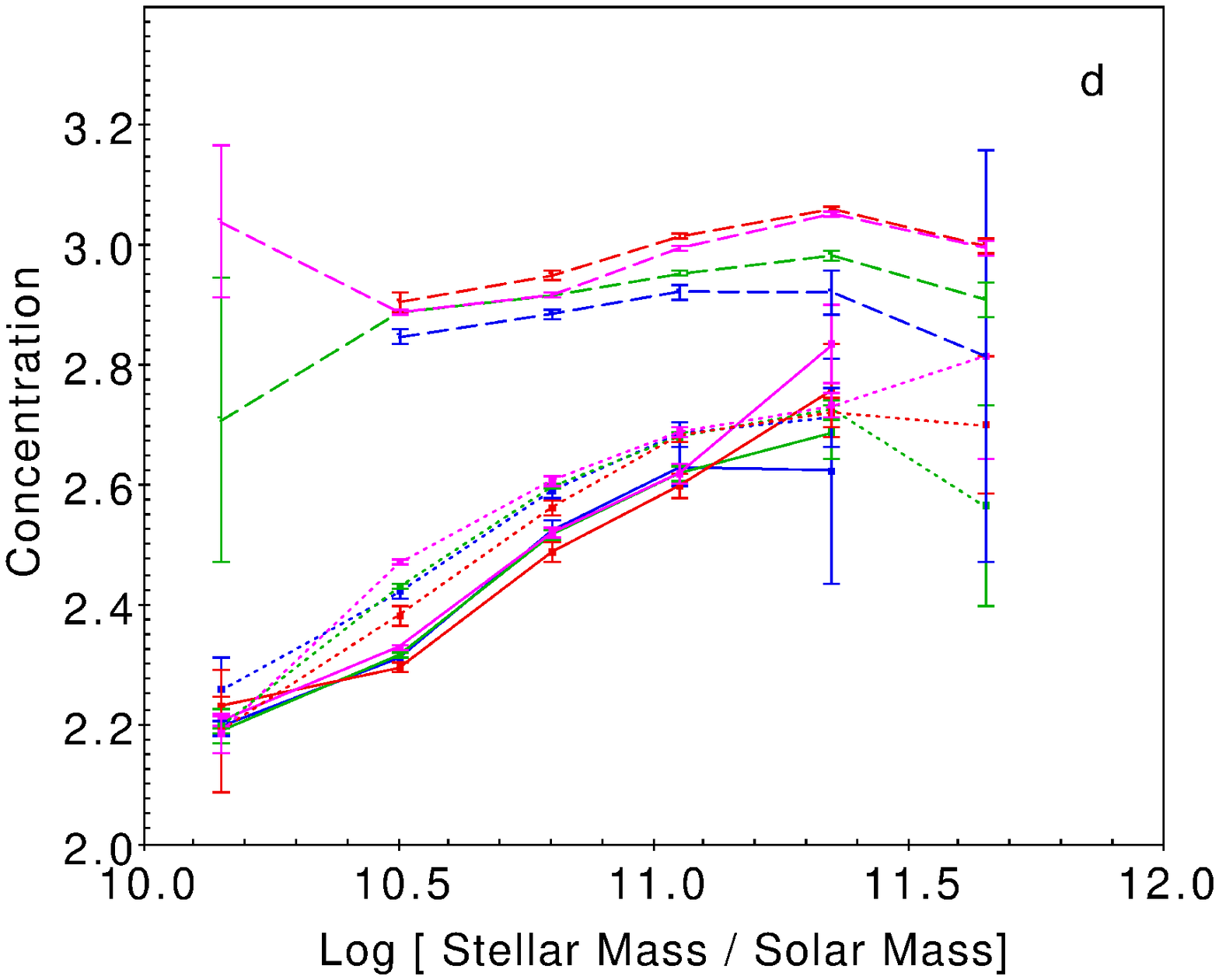}
 \includegraphics[ scale=0.35]{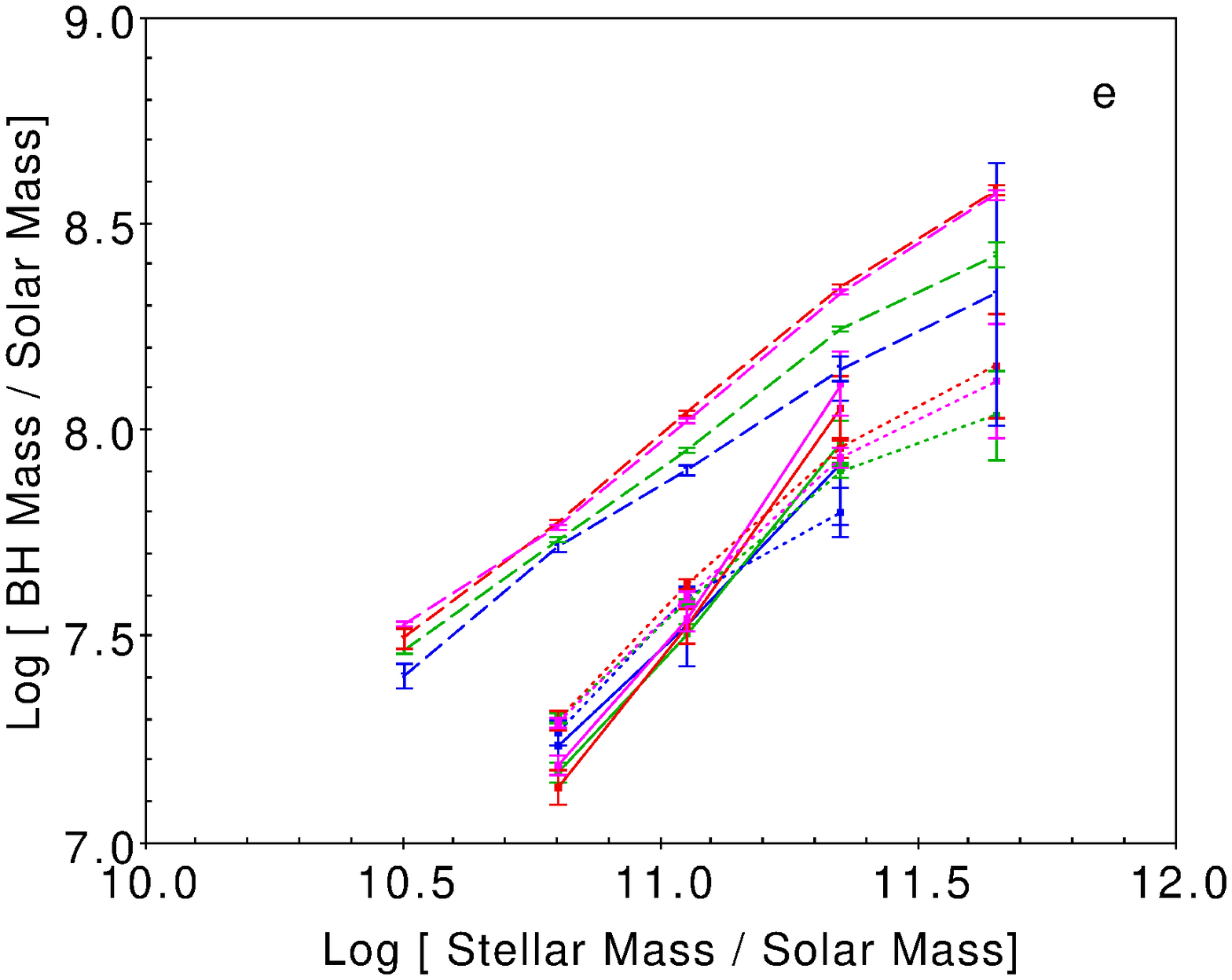}
 \includegraphics[ scale=0.35]{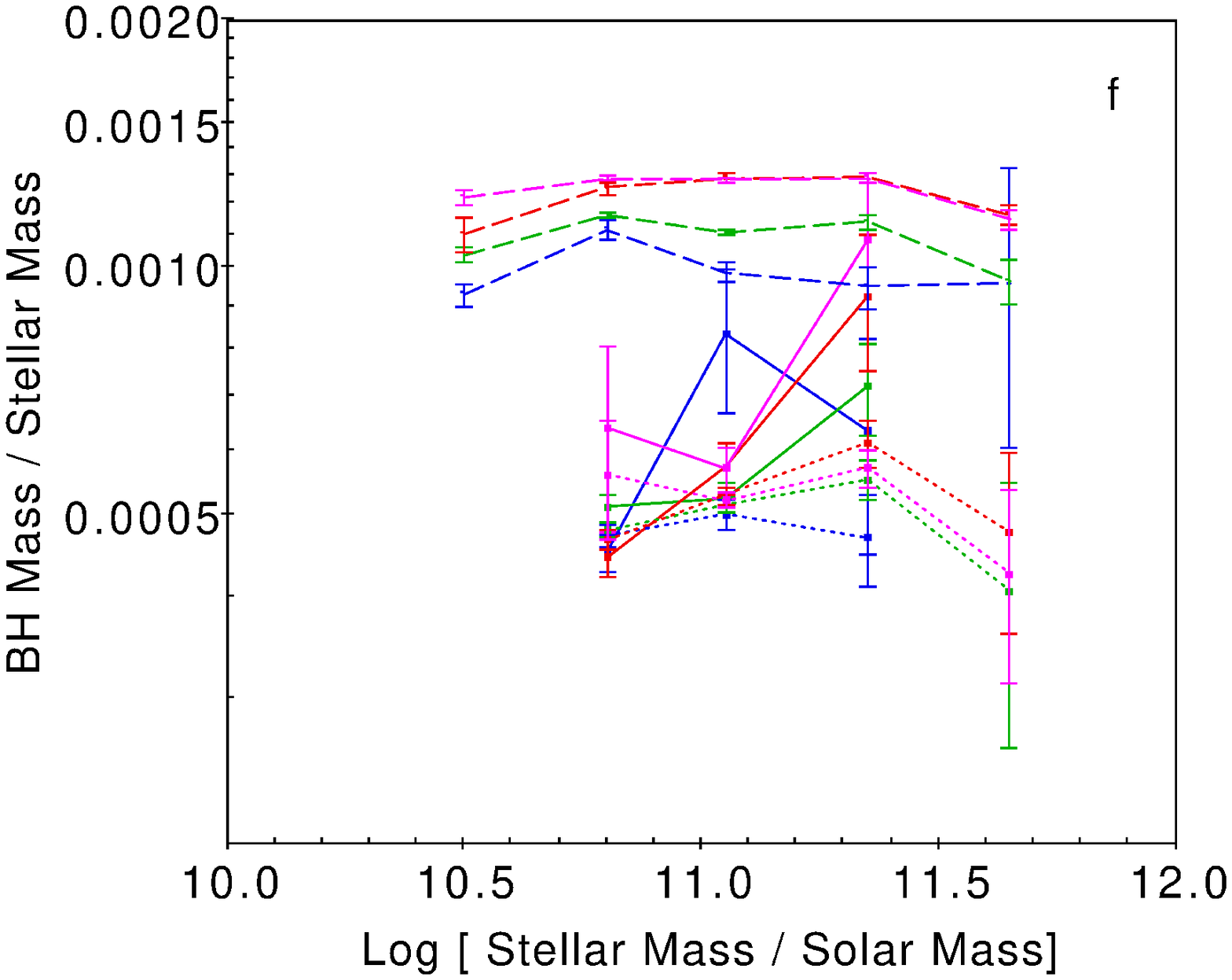}
 \includegraphics[ scale=0.35]{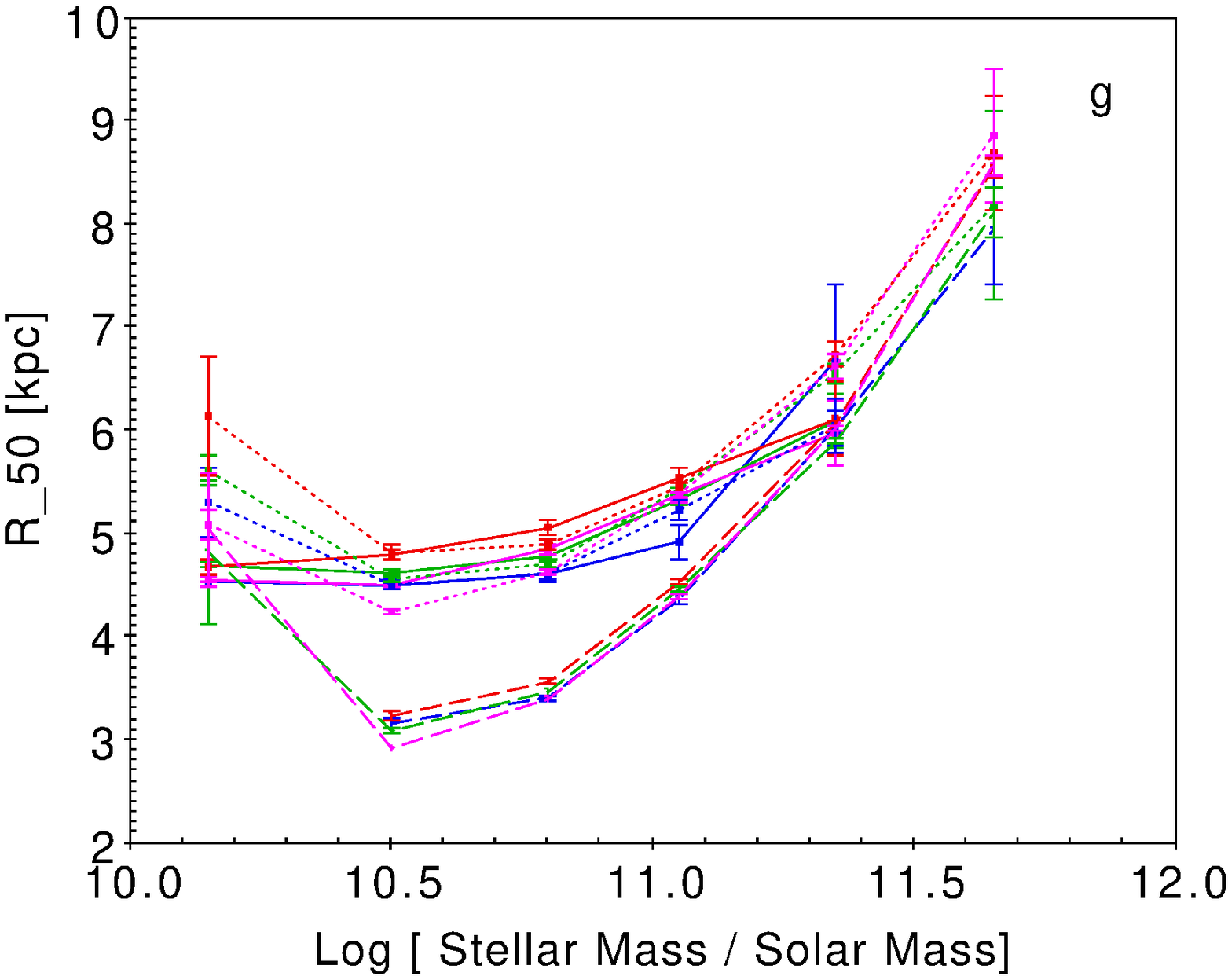}
 \includegraphics[ scale=0.35]{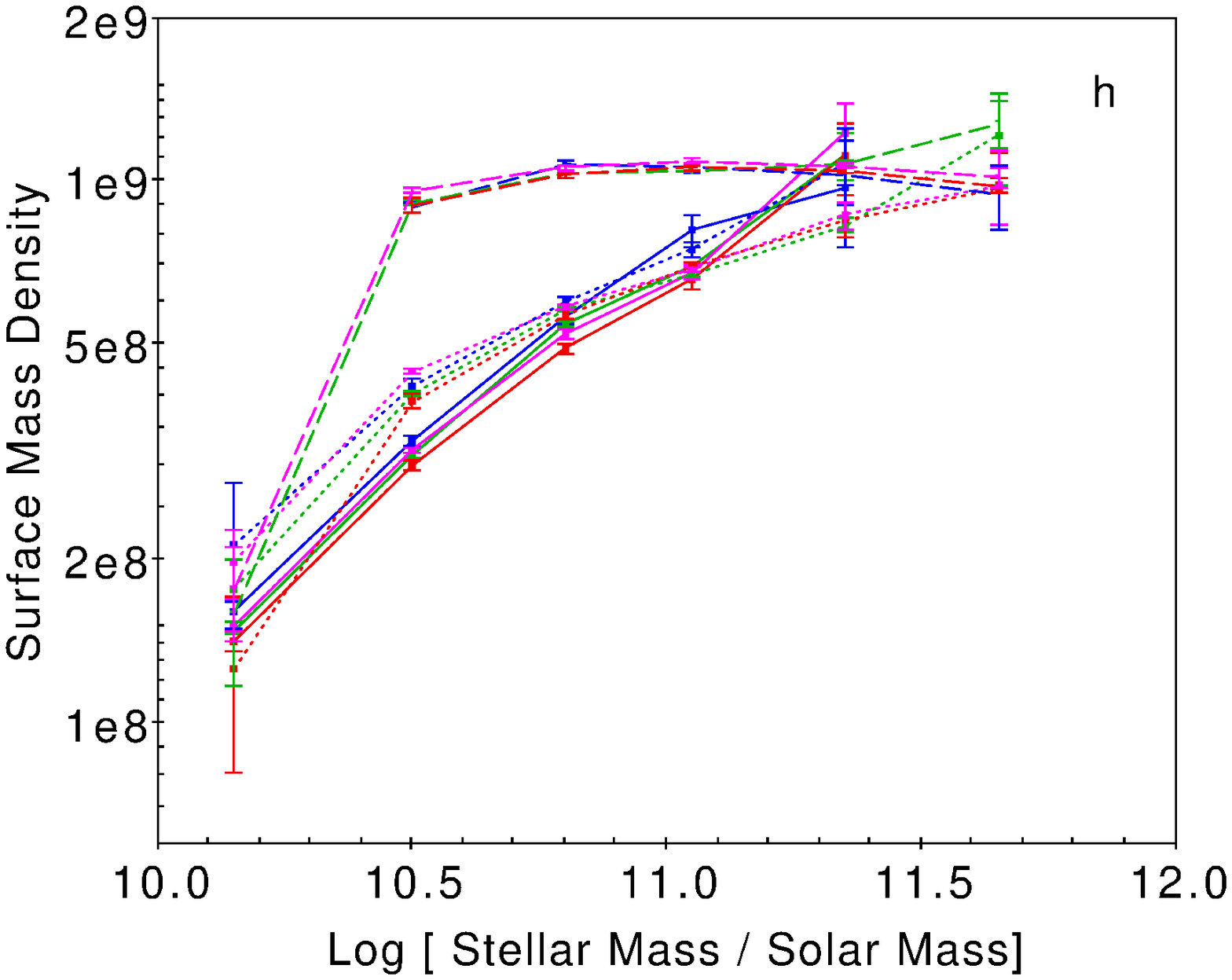}

\figcaption{The y-axis is the mean of each galaxy property. Colours represent galaxy samples with different environments
as in Fig. \ref{figophis}. Blue, green and red galaxies (as defined in Section
\ref{sec:colour}) are presented by the solid, dotted and dashed lines respectively. 
Error bars show the standard deviation of the mean.
\label{figopall}}
 \end{figure*}

The fraction of galaxies hosting optical and radio AGN in 
different stellar mass-colour bins are shown in Fig. \ref{figop}.
The AGN fraction depends strongly on the stellar mass in either optical or radio AGN.
The radio AGN activity shows a steeper dependence on 
the stellar mass than the optical AGN activity.
 The colour dependence is different in the two types of AGN activity.
The fraction of galaxies hosting radio-loud AGN decreases 
from red to blue galaxies while the opposite trend
is found for the optical AGN. The fraction of galaxies hosting 
optical AGN has a strong dependence on colour
which shows that optical AGN activity depends on the availability of 
cold and dense gas in galaxies (Heckman \& Best 2014).
The radio-mode AGN activity is higher in the red galaxies than in the blue and green galaxies
however there is an increase in the fraction of radio AGN for massive blue galaxies.
This result is consistent with the results of Janssen et~al.\ (2012) for the  
low excitation radio galaxies with the radio luminosity greater than 10$^{23}$  W Hz$^{-1}$.
The prevalence of radio-mode AGN activity in red galaxies is 
due to the abundant supply of hot gas from their surrounding environments to the SMBHs. The hot X-ray gas detected 
in galaxy groups and clusters, is the fueling source of radio AGN. Red galaxies, mostly found in 
dense environments such as galaxy groups and clusters, are expected to be fed by this gas. 
This will be investigated more in the following section. 

\begin{figure*}
\centering
 \includegraphics[ scale=0.25]{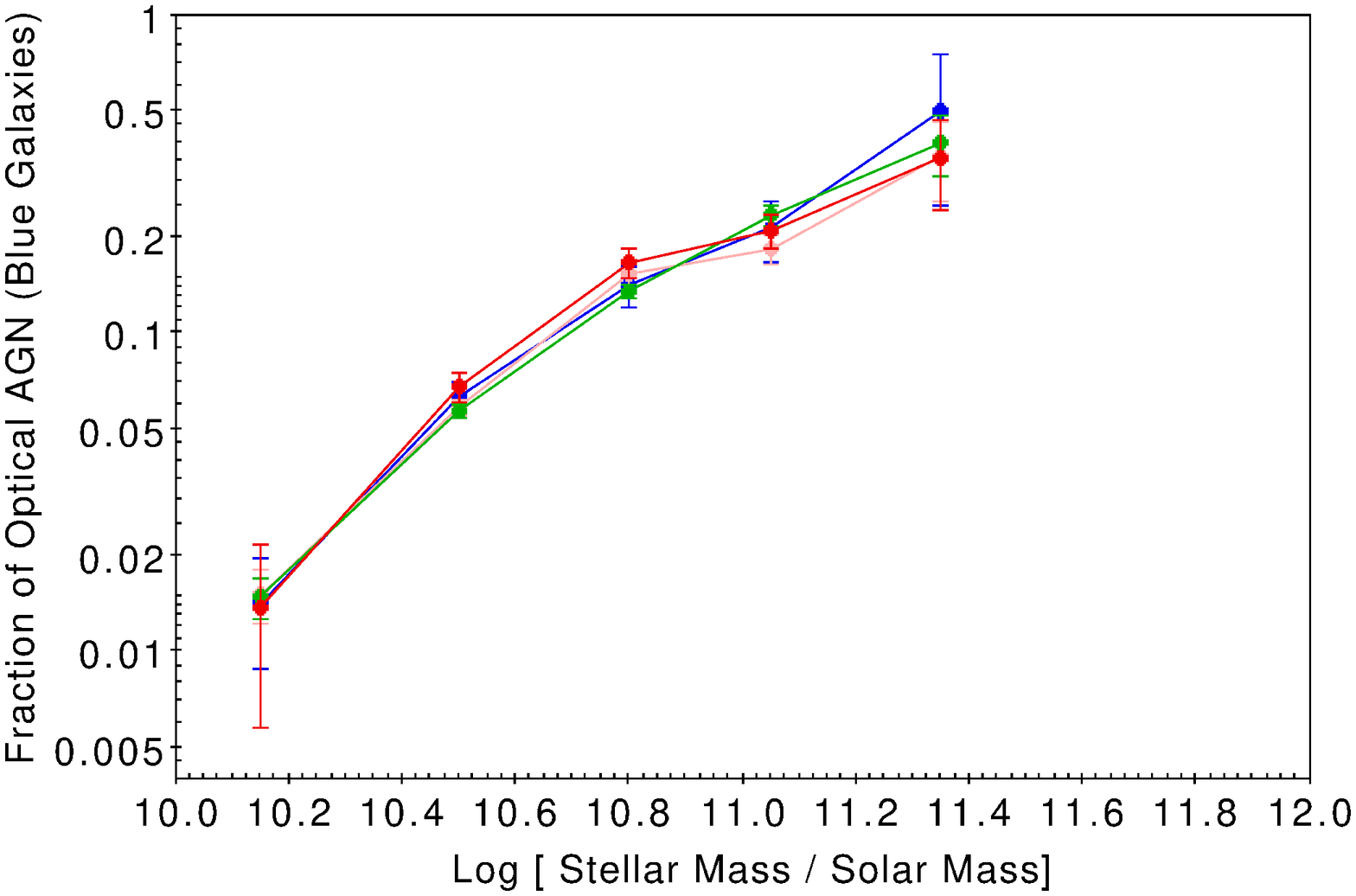}
 \includegraphics[ scale=0.25]{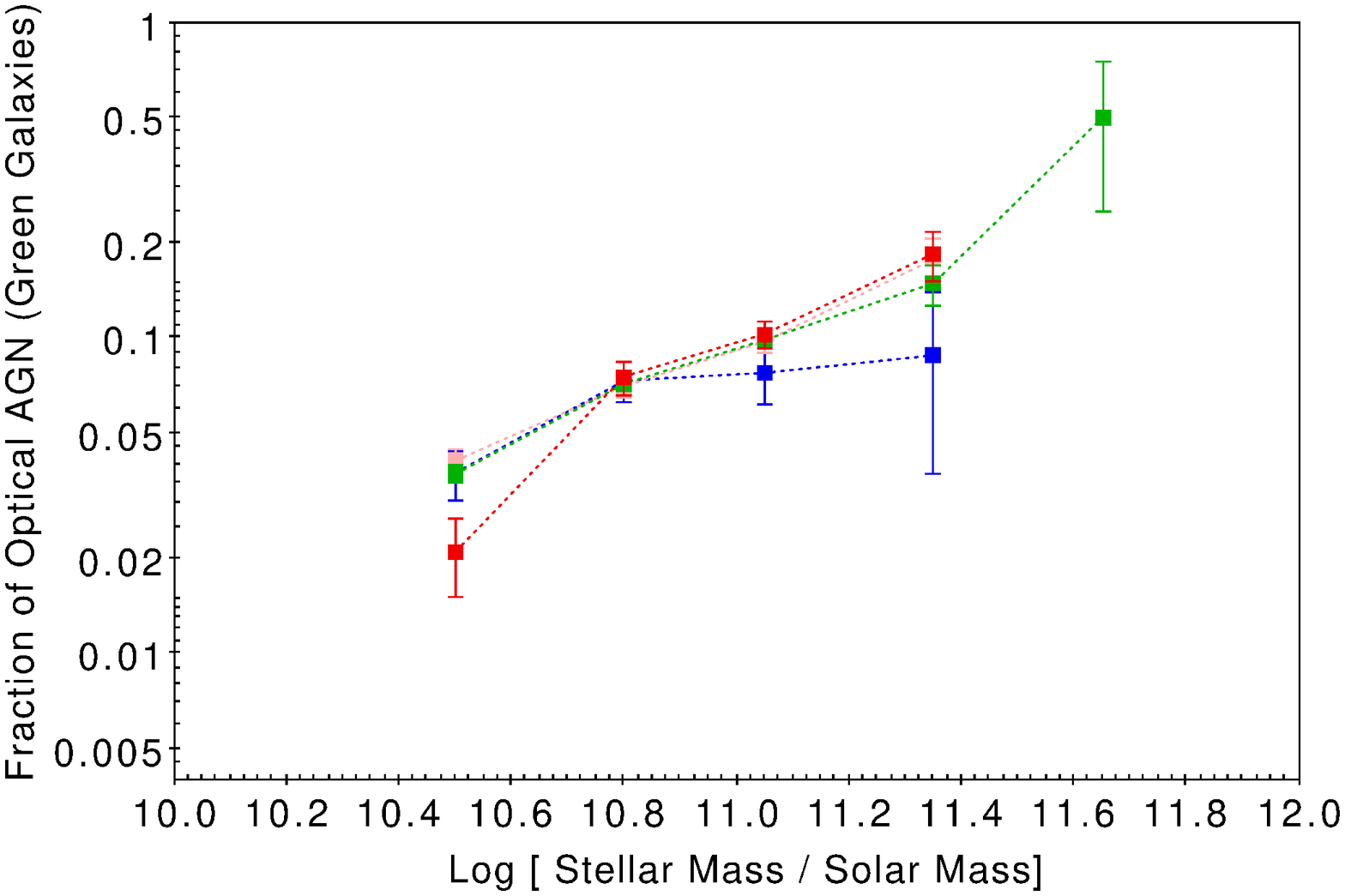}
 \includegraphics[ scale=0.25]{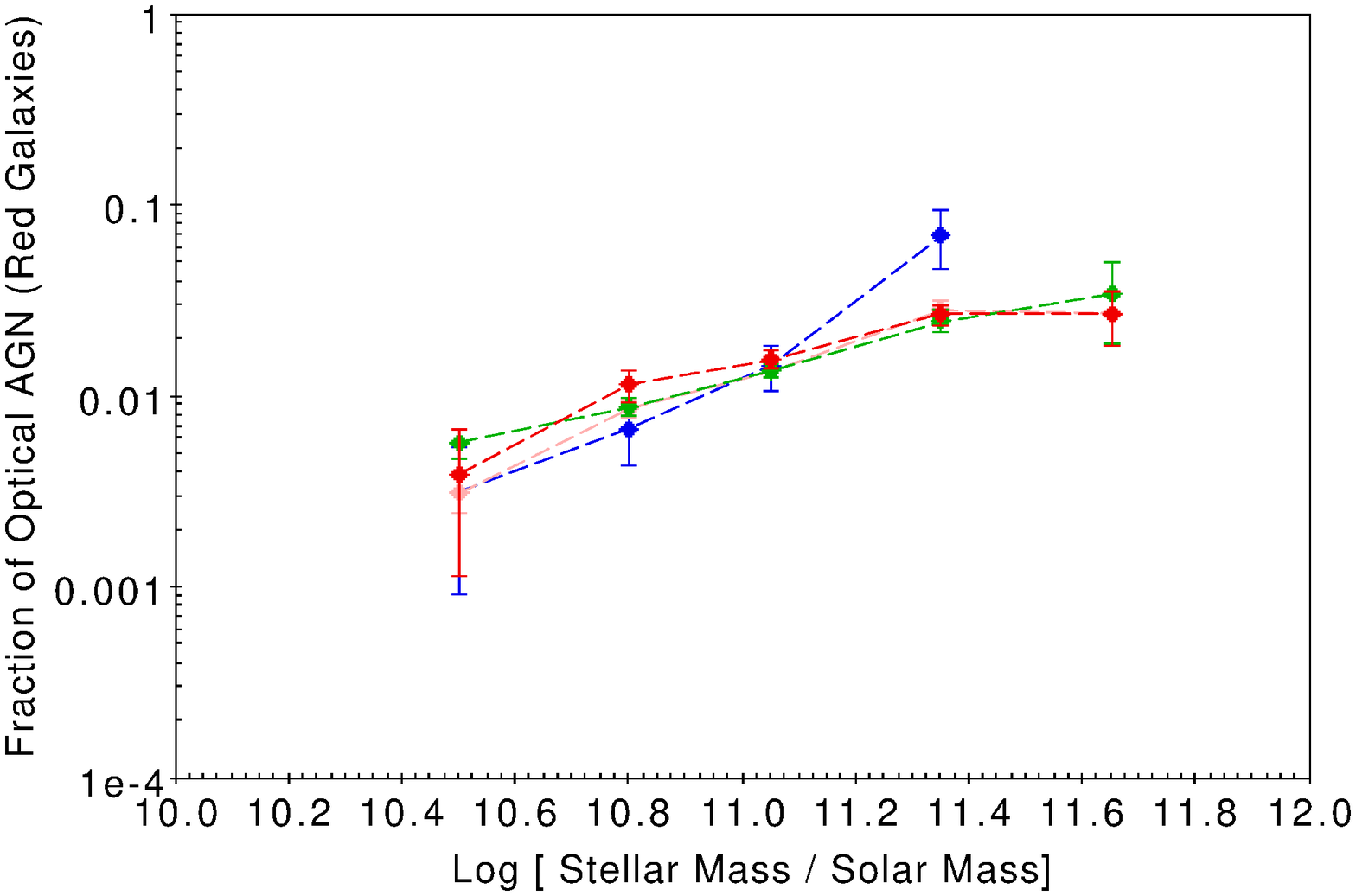}
 \includegraphics[ scale=0.25]{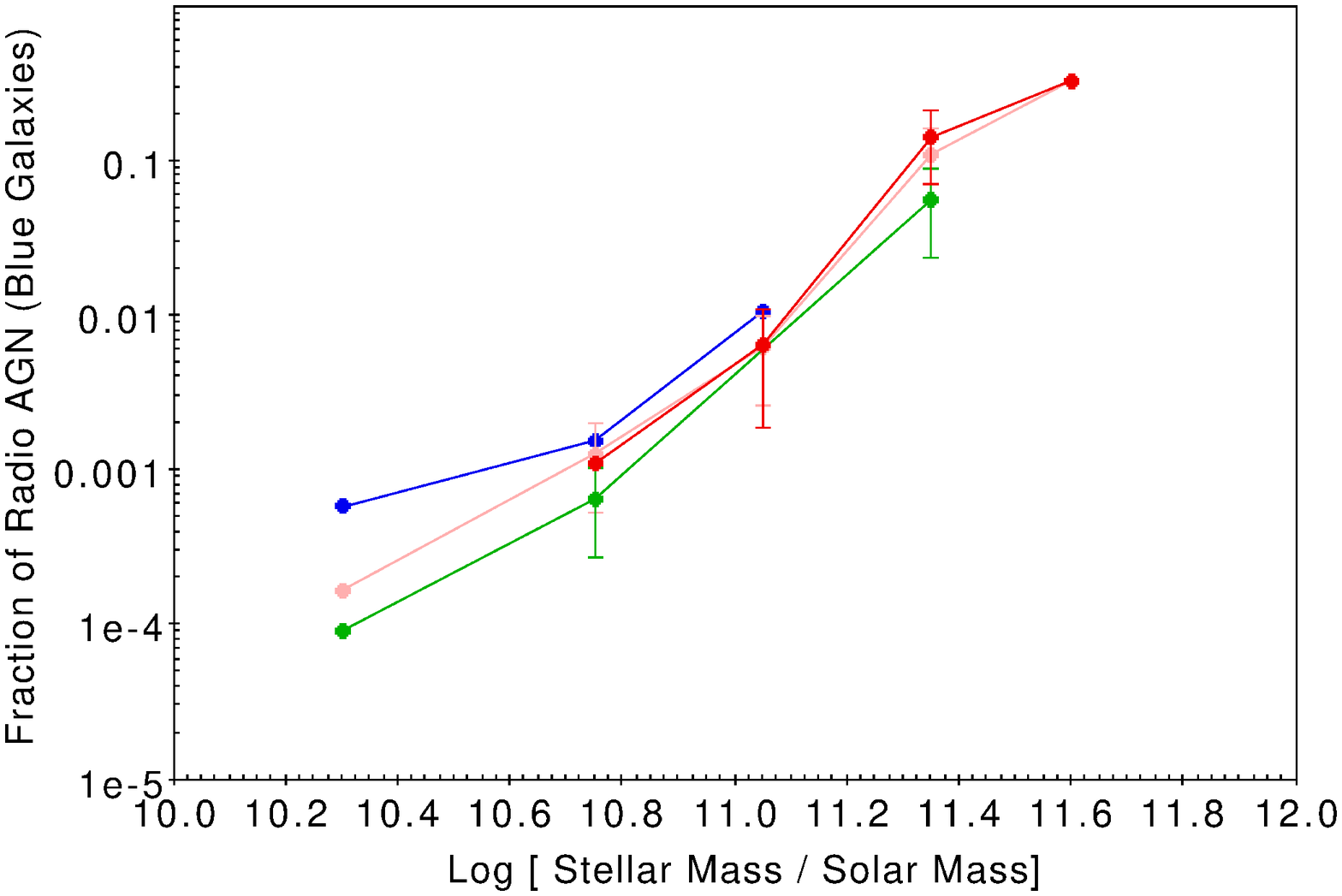}
 \includegraphics[ scale=0.25]{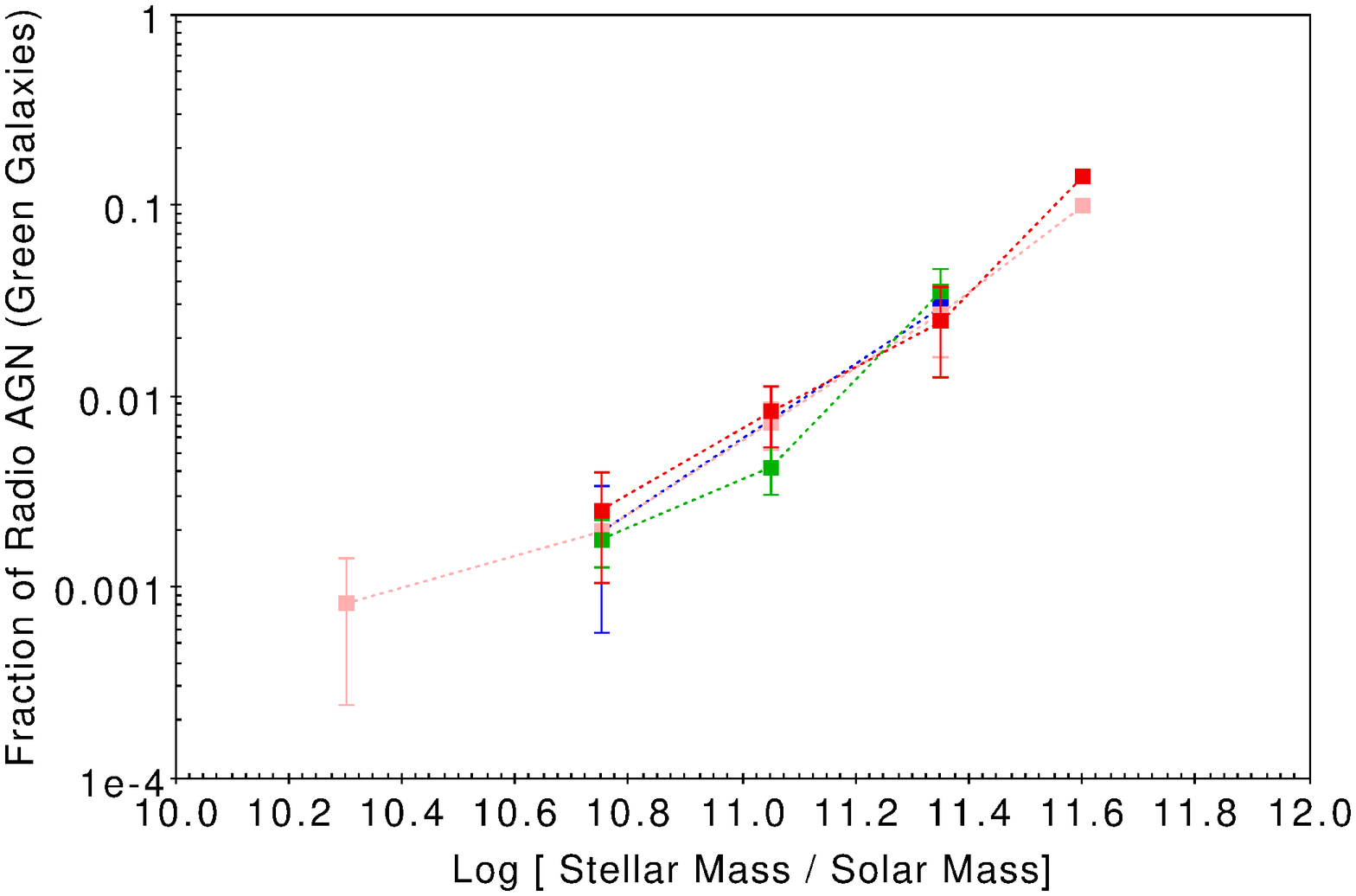}
 \includegraphics[ scale=0.25]{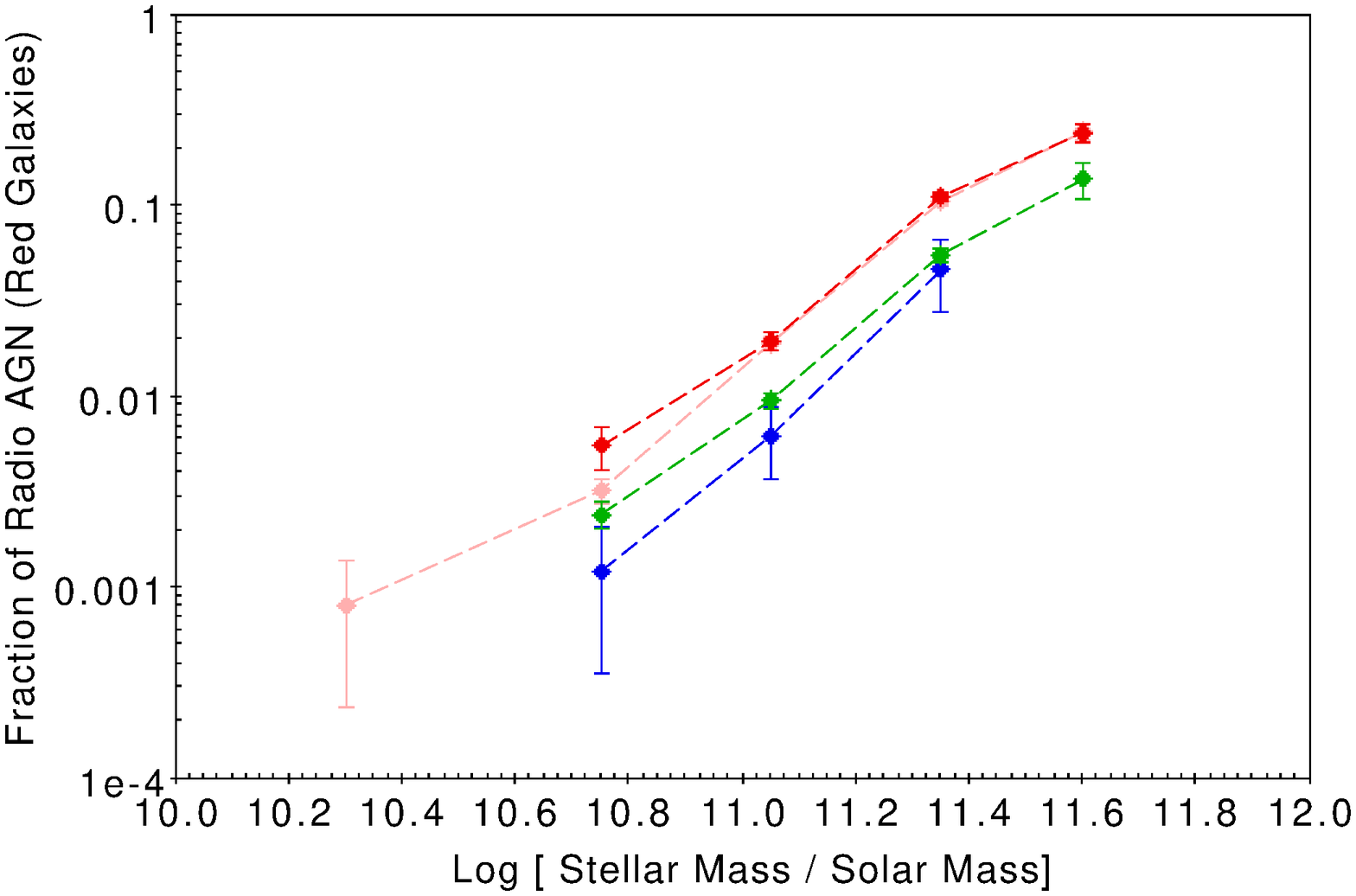}
\figcaption{Fraction of galaxies 
hosting optical (upper panels) and radio (lower panels) 
AGN in each stellar mass bin and for different environments. The colours represent 
galaxies in void (blue), isolated galaxies (green),
galaxies in group (pink) and BGGs (red) as in Fig. \ref{figophis}. The blue, green and red galaxies (as defined in Section
\ref{sec:colour}) are presented by the solid (left panels), dotted (middle panels) and dashed (right panels) lines respectively.
\label{figRGB}}
 \end{figure*}
  
In addition to the stellar mass and colour, the dependence of AGN activity on the black hole mass 
is also strong (Best  et~al.\ 2005b; Ishibashi et~al.\ 2014, {Bari{\v s}i{\'c}} et~al.\ 2017). 
By fixing the stellar mass and colour, the variations of other properties such as the black hole mass
have been investigated for galaxies of different environments (Fig. \ref{figopall}). 
This allows us to find out all the potential biases in the samples which may influence AGN activities.
The plots show the mean value of each parameter.
Blue, green and red galaxies are presented by solid, dotted and dashed lines respectively.
Galaxies in voids (blue), isolated galaxies (green), group galaxies (pink) and BGGs (red)
are shown in different colours. The error bars show the standard deviation of the mean.

There is a strong correlation between the black hole mass and the stellar mass 
in all types of galaxies (Fig.~5e). Fig.~5e omits the mean values of the black hole masses 
 for the stellar mass bin of 
M$_{\ast}$$\sim$10$^{10.2}$ \msun for the red, blue and green galaxies, as well as the stellar mass bin
of M$_{\ast}$$\sim$10$^{10.5}$ \msun for the blue and green galaxies. This is because
the number of objects with unreliable black hole masses (M$_{\rm BH}$$<$10$^{6.3}$ \msun) 
is significant in the low stellar mass bins, thus the mean values are uncertain.
In red galaxies, the concentration (Fig.~5d) and surface mass 
density (Fig.~5h) show only a little evolution as the stellar mass increases while these parameters significantly 
increased for blue and green galaxies.
It is well-illustrated by this plot that for blue (solid) and green (dotted) galaxies, if the stellar mass 
and colour are fixed, other host galaxy properties including the black hole mass do not change significantly
in response to the environment being changed. In red (dashed) galaxies, the void 
sample galaxies in the high stellar mass bins have on average lower
concentrations (Fig.~5d), lower black hole masses (Fig.~5e) and lower ratio of black hole to stellar mass ($\approx$3$\sigma$; Fig.~5f).
This may influence the results in Section \ref{sec:R}.
There are high scatters in the lowest and the highest mass bins 
due to the small number counts. Therefore, substantial care should be taken in interpreting the 
results in these bins.
The mean value of the stellar mass and 4000$\AA$ break in each stellar mass-colour bin
(not presented in Fig. \ref{figopall}) are matched well for all types of environments.

\section {AGN activity and galaxy environment  }
\label{sec:R}

In this section, AGN activity has been investigated for samples of galaxies 
with different environments and in fixed stellar mass-colour bins. 
The fraction of galaxies hosting AGN with respect to the stellar mass are plotted for galaxies 
of different colours and environments in Fig. \ref{figRGB}. The upper panels show optical AGN activity
and the lower panels show the radio-mode AGN activity. The blue, green 
and red galaxies are presented with solid, dotted and dashed lines respectively as in Fig. \ref{figopall}.
The colours indicate different environments as defined in Fig. \ref{figophis}.
The optical AGN activity in the blue galaxies does not show any dependence
on the environment. The green and red galaxy samples show the same results with
a lower S/N due to the smaller AGN sample size. The only exceptions 
are in the lowest mass bin for green BGGs and the highest mass bin for the red void
galaxies. For the former, it is worth noting that the BGGs in the lowest mass bin mostly belong to small
groups of two members corresponding to a very low halo mass and so the group identification uncertainty
is high in this stellar mass bin. Therefore, a possible explanation 
is a random error due to the small sample size in this bin. As it is shown in Fig.~\ref{figopall}d,
the mean of concentration is slightly lower for green BGGs 
in M$_{\ast}$$\sim$10$^{10.5}$ \msun, so this may also influence the observed difference there.
For the latter, in the red void galaxies in M$_{\ast}$$\sim$10$^{11.4}$ \msun, 
Fig.~\ref{figopall}d shows a significant decrease
in the concentration index of galaxies in voids compared to the galaxies in groups and in the BGG sample.
An increase in the fraction of optical AGN in the highest stellar mass bin is also seen
in Fig. \ref{figen} (right-hand panel), therefore the difference is not a random fluctuation.

\begin{figure*}
\centering
 \includegraphics[ scale=0.45]{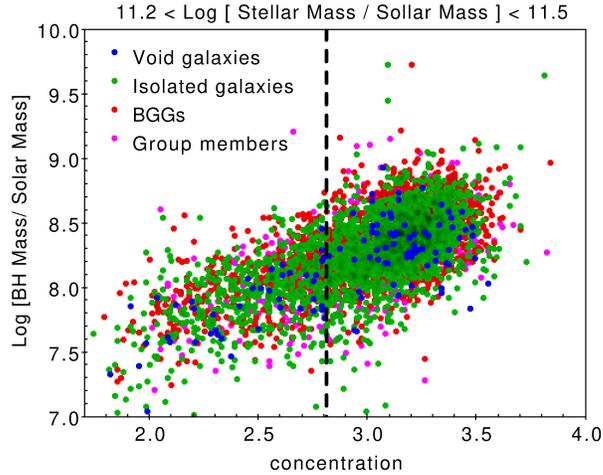}
\figcaption{Black hole mass with respect to the concentration index for the red galaxies 
in the stellar mass range 10$^{11.2-11.5}$ \msun. The colours represent different environments
 as in Fig. \ref{figophis}. Each sample has been divided into two subsamples based on the concentration
index. Division line of c=2.8 has been used here. Void galaxies in each concentration bins
show higher optical AGN fractions than the other three samples.
\label{figCon}}
 \end{figure*}

In order to investigate whether the bias from the concentration index causes 
this difference, optical AGN activity has been estimated for galaxy samples 
matched in concentration. Red galaxies with different environments in the stellar 
mass range M$_{\ast}$$\sim$10$^{11.2-11.5}$ \msun are considered.
The black hole mass versus the concentration index for these massive red galaxies
is plotted in Fig. 7. The colours represent different environments as in Fig. \ref{figophis}.
 There is a tail of low concentration void galaxies that have slightly low black hole masses.
To select galaxy samples matched in concentration, a division line C=2.8 is drawn in order to separate galaxies
into two bins of low and high concentrations. The optical AGN activities are then calculated
for each bin. The difference between the optical AGN activity of voids and the other three samples 
has been replicated for each concentration bin. 
Therefore, the higher fraction of optical AGN in massive void galaxies is not biased by the 
host galaxy properties. A higher fraction of optical AGN for galaxies in voids is also reported in 
Constantin et~al.\ (2007). In contrast, Amiri et~al.\ 2019 reported no enhancement in
the fraction of galaxies hosting optical AGN in voids compared to galaxy groups.

The radio-loud fractions in the blue, green and red galaxies are shown 
in the lower panels of Fig. \ref{figRGB}.
The highest S/N has been reached for the red galaxy samples.  The stellar mass-colour bins
with no error bars indicate there is only one radio source in the entire bin width. 
The fractions of radio-loud AGN in the red galaxies show strong dependence on the environment. 
Galaxies in voids as the most underdense regions have the lowest radio-loud fraction.
There is no difference between galaxies in groups and the BGGs in the stellar mass
bins greater than 10$^{11}$ M$_{\odot}$. 
The radio-loud fractions in blue and green galaxies show no dependence on the environment.
Although the number of void galaxies hosting radio AGN 
is not sufficient to be compared with the other galaxy samples,
the isolated and group galaxies and the BGGs with good 
statistics show that there is no significant difference
between radio-mode AGN activity in different
environments for either blue (solid) or green (dotted) galaxies.
Larger radio-loud samples of blue and green galaxies especially in voids are required to
reach a robust conclusion. The radio-mode AGN activity increases 
in dense environments in red elliptical galaxies where the hot IGM gas seems to be the 
only feeding source while the role of environment is not significant 
 in blue and green galaxies in the presence
of cold gas as a dominant source of feeding for the SMBH.

Combining the results of optical and radio AGN shows that 
environment does not contribute greatly to fueling the SMBH in blue and
green galaxies. The effect of environment on AGN activity
can be better detected in the absence of cold gas in the red quenched galaxies.
In the radio-mode AGN activity which hot accretion dominates, 
fueling of AGN from the hot IGM gas may decrease in the presence of
cold gas. In other words, the efficiency of
hot accretion depends on the host galaxy properties. 
In optical AGN activity, the SMBHs are assumed to be fed mainly by 
the star forming cold gas that is available inside the galaxies
or brought through the filaments in the local environment of the galaxies. 
Therefore, the effect of environment can be better seen
in the red galaxies where usually there is no or little amount of cold gas.
In this regard, it is shown that massive void galaxies have a significantly
higher fraction of optical AGN. This can be considered as direct 
evidence for the effect of environment on optical AGN
activity as void galaxies experience a higher level of one-on-one interaction which 
 is trivially more pronounced in the massive galaxies. A more fundamental 
reason related to the formation and evolution of galaxies in voids may also be responsible for this, 
which indirectly confirms the effect of environment on optical AGN activity.

\section{Summary and conclusion}
\label{sec:summery}

In this study, the radio and optical AGN activities of galaxies in different
environments have been investigated. Samples of galaxies in groups, BGGs and isolated
galaxies from Tago et~al.\ (2010), along with a sample of galaxies in voids from 
Pan et~al.\ (2012) have been used for this aim. The combined radio and optical data of galaxies
including the SDSS value-added spectroscopic 
catalogue adapted from Best \& Heckman (2012), are also used to extract the 
galaxy properties. The environmental parameters have been drawn from Sabater et~al.\ (2013).

The galaxy samples display different host galaxy and environmental properties.
The void sample have a higher fraction of late-type and blue star-forming galaxies 
than the other three samples. The BGG and galaxy group samples have higher
fractions of early-type, red and quenched galaxies than the isolated galaxy and 
void galaxy samples. The environmental parameters show that galaxies in voids have
the lowest-density environment and the isolated galaxies, group galaxies
and the BGGs show increasingly higher-density environments, in that order. In terms of one-on-one
interaction, it is shown that galaxies in voids have the highest level of interaction.

In order to find out the effect of environment on AGN activity, the biases 
caused by the host galaxy properties have been removed. Both the radio-mode and optical
AGN activities show a strong dependence on the stellar mass and colour of the host galaxy.
Using galaxy samples matched in stellar mass and colour, the optical and radio-mode
AGN activities have been investigated in different environments.
No dependence on the environment is found in either optical or radio AGN activity in blue
galaxies nor in green galaxies with lower S/N. The effect of environment on AGN activity
is significant in red galaxies. This implies that the efficiency of gas accretion  
from the close or large-scale environments into the SMBHs
 is low in the presence of cold gas in the galaxies. 
Red galaxies in dense environments such as the BGGs and group member galaxies 
have higher fractions of radio-loud AGN compared to the galaxies in voids
and isolated galaxies. Therefore, in the absence of fresh cold 
gas in the galaxy, the hot IGM gas in dense environment
efficiently triggers radio-mode AGN activity.
Massive red galaxies in voids show a higher optical AGN fraction than
other galaxy samples. A more detailed investigation was performed to show that
in addition to the stellar mass and colour, this result is not biased by 
the concentration and black hole mass of the host galaxies. 
The results provide new evidence for the effect of environment on optical 
AGN activity which may be due to the higher level of one-on-one interaction 
in void galaxies or it may be derived from a more fundamental difference in the
formation and evolution of galaxies in voids compared to the denser environments.


\begin{thebibliography}{}
\bibitem[Abazajian et al. 2009]{Abazajian09} 
Abazajian, K. N., 
Adelman-McCarthy, J. K., Agueros, M. A., et~al.\ 2009,
ApJS, 182, 543
\bibitem[Amiri et al. 2019]{Amiri19}
Amiri, A., Tavasoli, S., \& De Zotti, G. 2019, ApJ, 874, 140
\bibitem[Argudo-Fernandez et al. 2018]{Argudo18}
{{Argudo-Fern{\'a}ndez}}, M., Lacerna, I., \& Duarte Puertas, S. 2018, 
A\&A, 620, 113 
\bibitem[Argudo-Fernandez et al. 2016]{Argudo16}
Argudo-Fern{\'a}ndez, M., Shen, S., Sabater, J., et~al.\ 2016, A\&A,
592, 30
\bibitem[Baldry et al. 2006]{Baldry06}
Baldry, I. K., Balogh, M. L., Bower, R. G., et~al.\ 2006, MNRAS, 373, 469
\bibitem[Baldwin et al. 1981]{Baldwin81}	
Baldwin, J. A., Phillips, M. M., \& Terlevich, R. 1981, PASP, 93, 5
\bibitem[Balogh et al. 2004]{Balogh04}
Balogh, M., Eke V., Miller, C., et~al.\ 2004, MNRAS, 348, 1355
\bibitem[Barisic et al. 2017]{Barisic17}
{{Bari{\v s}i{\'c}}}, I., {van der Wel}, A., {Bezanson}, R., et~al.\
 2017, MNRAS, 847, 9
\bibitem[Becker et al. 1995]{Becker95}
Becker, R. H., White, R. L., \& Helfand, D. J. 1995, ApJ, 450, 559
\bibitem[Best \& Heckman 2012]{Best12}
Best, P. N., \& Heckman, T. M. 2012, MNRAS, 421, 1569
\bibitem[Best et al. 2005a]{Best05a}	
Best, P. N., Kauffmann, G., Heckman, T. M., \& Ivezic, Z. 2005a, MNRAS, 362, 9
\bibitem[Best et al. 2005b]{Best05b}
Best, P. N., Kauffmann, G., Heckman, T. M., et~al.\ 2005b, MNRAS, 362, 25
\bibitem[Best et al. 2007]{Best07}
Best, P. N., von der Linden, A., Kauffmann, G., Heckman, T. M., \& Kaiser C. R.
2007, MNRAS, 379, 894
\bibitem[Bradshaw et al. 2011]{Bradshaw11}
Bradshaw, E. J., Almaini, O., Hartley, W. G., et~al.\ 2011, MNRAS, 415, 2626
\bibitem[Brinchmann et al. 2004]{Brinchmann04}
Brinchmann, J., Charlot, S., White, S. D. M., et~al.\ 2004, MNRAS, 351, 1151
\bibitem[{Condon et al. }{1998}]{Condon98}
Condon, J. J., Cotton, W. D., Greisen, E. W.,  et~al.\
1998, AJ, 115, 1693
\bibitem[{Deng et al. }{2012}]{Deng12} 
Deng, X., Song, J., Chen, Y., Jiang, P., \& Ding, Y. 2012,
ApJ, 753, 166
\bibitem[{Garofalo et al. }{2010}]{Garofalo10} 
Garofalo, D., Evans, D. A., \& Sambruna, R. M. 2010, MNRAS, 406, 975
\bibitem[{Gilmour et al. }{2007}]{Gilmour07} 
Gilmour, R., Gray, M. E., Almaini, O., et al. 2007, MNRAS, 380, 1467
\bibitem[{Gilmour et al. }{2018}]{Gilmour18} 
Gordon, Y. A., Pimbblet, K. A., Owers, M. S.,  et~al.\ 2018, MNRAS, 475, 4223 
\bibitem[{Heckman \&  Best }{2014}]{Heckman14}
Heckman, T. M., \& Best, P. N. 2014, ARA\&A, 52, 589
\bibitem[{Hong et al.}{2015}]{Hong15}
Hong, J., Im, M., Kim, M., \& Ho, L. C. 2015,
ApJ, 804, 34
\bibitem[{Hoyle et al.}{2005}]{Hoyle05}
Hoyle, F., Rojas, R. R., Vogeley, M. S., \& Brinkmann, J. 2005,
ApJ, 620, 618
\bibitem[{Hwang et al.}{2012}]{Hwang12}
Hwang, H. S., Park, C., Elbaz, D., \& Choi, Y. -Y. 2012, A\&A, 538, 15
\bibitem[{Ishibashi  et al. }{2014}]{Ishibashi14}
Ishibashi, W., Auger, M. W., Zhang, D., \& Fabian, A. C. 2014, MNRAS,
443, 1339
\bibitem[{Janssen et al. }{2012}]{Janssen12}
Janssen, R. M. J., {R{$\ddot{o}$}ttgering}, H. J. A., Best, P. N., \& Brinchmann J.
2012, A\&A, 541, 62
\bibitem[{Kauffmann  et al. }{2003}]{Kauffmann03a}
Kauffmann, G., Heckman, T. M., Tremonti, C., et al. 2003a, MNRAS, 346, 1055
\bibitem[{Kauffmann  et al. }{2003}]{Kauffmann03b}
Kauffmann, G., Heckman, T. M., White, S. D. M., et al. 2003b, MNRAS, 341, 33
\bibitem[{Kauffmann et al.}{2004}]{Kauffmann04}	
Kauffmann, G., White, S. D. M., Heckman, T. M., et~al.\ 2004, MNRAS, 353, 713
\bibitem[{Kewley et al. }{2006}]{Kewley06}
Kewley, L. J., Groves, B., Kauffmann, G., \& Heckman, T. 2006, MNRAS, 372, 961
\bibitem[{Khosroshahi et al. }{2017}]{Khosroshahi17}
Khosroshahi, H. G., Raouf, M., Miraghaei, H., et~al.\ 2017, 
ApJ, 842, 81 
\bibitem[{Kolmogorov }{1933}]{Kolmogorov1933}
Kolmogorov, A. 1933, Inst. Ital. Attuari, Giorn., 4, 83.
\bibitem[{Koulouridis et al. }{2018}]{Koulouridis18}
Koulouridis, E., Ricci, M., Giles, P., et~al.\ 2018, A\&A, 620, 20
\bibitem[{Kreckel et al.}{2011}]{Kreckel11}
Kreckel, K., Platen, E., Aragon-Calvo, M. A., et~al.\ 2011, AJ, 141, 4
\bibitem[{Kreckel et al.}{2012}]{Kreckel12}
Kreckel, K., Platen, E., Aragon-Calvo, M. A., et~al.\ 2012, AJ, 144, 16
\bibitem[{Kreckel et al.}{2014}]{Kreckel14}
Kreckel, K., van Gorkom, J. H., Beygu, B., et~al.\ 2014, in Proceedings of IAU Symposium 308 
"The Zeldovich Universe: Genesis and Growth of the Cosmic Web", arXiv:1410.6597
\bibitem[{Li et al.}{2019}]{Li19}
Li, F., Gu, Y.-Z., Yuan, Q.-R., et~al.\ 2019, MNRAS, 484, 3806 
\bibitem[{Lopes et al.}{2017}]{Lopes17}
Lopes, P. A. A., Ribeiro, A. L. B., \& Rembold, S. B. 2017, MNRAS, 472, 409
\bibitem[{Magliocchetti et al. }{2018}]{Magliocchetti18}
Magliocchetti, M., Popesso, P., Brusa, M., \& Salvato, M. 2018,
MNRAS, 478, 3848
\bibitem[{Malavasi et al. }{2015}]{Malavasi15}
Malavasi, N., Bardelli, S., Ciliegi, P., et~al.\ 2015,  
A\&A, 576, 101
\bibitem[{Man et al.}{2019}]{Man19} 
Man, Zh., Peng, Y., Kong X., et~al.\ 2019, MNRAS, 488, 89
\bibitem[{Manzer \& De Robertis}{2014}]{Manzer14}
Manzer, L. H., \& De Robertis, M. M. 2014, 788, 140
\bibitem[{McNamara et al. }{2011}]{McNamara11}
McNamara, B. R., Rohanizadegan, M., \& Nulsen, P. E. J. 2011, ApJ, 727, 39
\bibitem[{Miller et al.}{2003}]{Miller03} 
Miller, C. J., Nichol, R. C., G{\'o}mez, P. L., Hopkins, A. M., \& Bernardi, M. 2003,
ApJ, 597, 142
\bibitem[{Miraghaei \& Best}{2017}]{Miraghaei17}
Miraghaei, H., \& Best, P. N. 2017, MNRAS, 466, 4346
\bibitem[{Miraghaei et al.}{2014}]{Miraghaei14} 
Miraghaei, H., Khosroshahi, H. G., Kl{$\ddot{o}$}ckner, H. -R.,
et~al.\ 2014, MNRAS, 444, 651
\bibitem[{Miraghaei et al.}{2015}]{Miraghaei15}
Miraghaei, H., Khosroshahi, H. G., Sengupta, C., et~al.\ 2015, AJ, 150, 196
\bibitem[{Pan et al.}{2012}]{Pan12}
Pan, D. C., Vogeley, M. S., Hoyle, F., Choi, Y., \& Park, C. 2012, MNRAS, 421, 926 
\bibitem[{Pimbblet et al.}{2013}]{Pimbblet13}
Pimbblet, K. A., Shabala, S. S., Haines, C. P., Fraser-McKelvie, A., \& Floyd, D. J. E.
2013, MNRAS, 429, 1827
\bibitem[{Ricciardelli et al. }{2017}]{Ricciardelli17}  
Ricciardelli, E., Cava, A., Varela, J., \& Tamone, A. 2017,
ApJL, 846, L4
\bibitem[{Rojas et al. }{2005}]{Rojas05} 
Rojas, R. R., Vogeley, M. S., Hoyle, F., \& Brinkmann, J. 2005,
ApJ, 617, 50
\bibitem[{Sabater et al. }{2013}]{Sabater13} 
Sabater, J., Best, P. N., \& Argudo-Fernandez, M. 2013, MNRAS, 430, 638
\bibitem[{Sabater et al. }{2019}]{Sabater19} 
Sabater, J., Best, P. N., Hardcastle, M. J., et~al.\ 2019, A\&A, 622, 17 
\bibitem[{Sabater et al. }{2015}]{Sabater15} 
Sabater J., Best P. N., \& Heckman T. M. 2015, MNRAS, 447, 110
\bibitem[{Sabater et al. }{2012}]{Sabater12} 
Sabater, J., Verdes-Montenegro, L., Leon, S., Best, P., \& Sulentic J.
2012, A\&A, 545, 15
\bibitem[{Strauss et al. }{2002}]{Strauss02}
Strauss, M. A., Weinberg, D. H., Lupton, R. H., et al. 2002, AJ, 124, 1810	
\bibitem[{Tago et al. }{2010}]{Tago10}	
Tago, E., Saar, E., Tempel, E., et~al.\ 2010
A\&A, 514, 102
\bibitem[{Tremaine et al. }{2002}]{Tremaine02}	
Tremaine, S., Gebhardt, K., Bender, R., et al. 2002, ApJ, 574, 740
\bibitem[{Von Der Linden et al.}{2007}]{Linden07}
Von Der Linden, A., Best, P. N., Kauffmann, G., \& White, S. D. M. 2007, MNRAS,
379, 867
\bibitem[{Wang \& Li}{2019}]{Wang19}
Wang, L., \& Li, C. 2019, MNRAS, 483, 1452 
\bibitem[{York et al. }{2000}]{York00}
York, D. G., Adelman, J., Anderson, J. E., et al. 2000, AJ, 120, 1579


\end{thebibliography}
\end{document}